# Practical Guidelines for Cell Segmentation Models Under Optical Aberrations in Microscopy

Boyuan Peng[1,2,3#], Jiaju Chen[1,2,3#], P. Bilha Githinji[1,2,3#], Ijaz Gul[1,2,3], Qihui Ye[1,2,3], Minjiang Chen[1], Peiwu Qin[1,2,3], Xingru Huang[3], Chenggang Yan[3], Dongmei Yu[3,4]*, Jiansong Ji[1]*, Zhenglin Chen[1,2,3]*

1. Zhejiang Key Laboratory of Imaging and Interventional Medicine, Zhejiang Engineering Research Center of Interventional Medicine Engineering and Biotechnology, The Fifth Affiliated Hospital of Wenzhou Medical University, Lishui 323000, China
2. Institute of Biopharmaceutical and Health Engineering, Shenzhen International Graduate School, Tsinghua University, Shenzhen, Guangdong, China
3. School of Automation, Hangzhou Dianzi University, Hangzhou, Zhejiang Province, 310018, China
4. School of Mechanical, Electrical & Information Engineering, Shandong University, Weihai, Shandong 264209, China

#These authors contributed equally to this work.
*Corresponding authors. E-mail: yudongmei198011@sina.cn (P. Yu); jjstcty@sina.com (P. Ji); zhenglin.chen@sz.tsinghua.edu.cn (D. Chen)

**Abstract**

Cell segmentation is essential in biomedical research for analyzing cellular morphology and behavior. Deep learning methods, particularly convolutional neural networks (CNNs), have revolutionized cell segmentation by extracting intricate features from images. However, the robustness of these methods under microscope optical aberrations remains a critical challenge. This study evaluates cell image segmentation models under optical aberrations from fluorescence and bright field microscopy. By simulating different types of aberrations, including astigmatism, coma, spherical aberration, trefoil, and mixed aberrations, we conduct a thorough evaluation of various cell instance segmentation models using the DynamicNuclearNet (DNN) and LIVECell datasets, representing fluorescence and bright field microscopy cell datasets, respectively. We train and test several segmentation models, including the Otsu threshold method and Mask R-CNN with different network heads (FPN, C3) and backbones (ResNet, VGG, Swin Transformer), under aberrated conditions. Additionally, we provide usage recommendations for the Cellpose 2.0 Toolbox on complex cell degradation images. The results indicate that the combination of FPN and SwinS demonstrates superior robustness in handling simple cell images affected by minor aberrations. In contrast, Cellpose 2.0 proves effective for complex cell images under similar conditions. Furthermore, we innovatively propose the Point Spread Function Image Label Classification Model (PLCM). This model can quickly and accurately identify aberration types and amplitudes from PSF images, assisting researchers without optical training. Through PLCM, researchers can better apply our proposed cell segmentation guidelines. This study aims to provide guidance for the effective utilization of cell segmentation models in the presence of minor optical aberrations and pave the way for future research directions.

**Introduction**

Cell imaging plays a critical role in the biological sciences, enabling researchers to visualize and analyze the complex structures and functions of cells with high precision. Among the various imaging modalities available, fluorescence microscopy and bright field microscopy are the most widely used techniques. Fluorescent microscopes and bright field microscopes are both essential techniques in cell imaging, each with distinct advantages and limitations. The bright field microscopes are the most popular and widely used of all microscopes [1] due to their simplicity and the ability to use unstained images, which preserves cell integrity and allows for morphology-based classification without the complications of fluorescent markers, such as cytotoxicity and nonspecific labeling [2]. One of the main applications for bright field microscopy is classifying and counting distinct cell lineages, including SH-SY5Y, Huh7, and A549 cells, demonstrating its versatility in imaging different cell types [3]. Furthermore, high-contrast brightfield microscopy was specifically used to count HEK293 and primary human dermal fibroblast cells, showcasing its effectiveness in noninvasive cell proliferation assays [4]. Conversely, fluorescent microscopes excel in visualizing specific cellular structures through differential staining, which is crucial for detailed cellular analysis. Such advantages make fluorescence microscopes play a vital role in cell nucleus research, providing a variety of advanced imaging features essential for cell nucleus research. For instance, fluorescence microscopy allows researchers to explore the architecture of the cell nucleus, revealing how dynamic structures such as chromatin and nuclear bodies influence protein dynamics and interactions within the nucleoplasm. Additionally, fluorescence microscopy enables the observation of molecular interactions in real-time, offering a powerful approach to studying protein localization and interactions under physiological conditions. Whether using bright field microscopy for sorting and counting cell lineages or fluorescence microscopy for nuclear physiological studies, cell segmentation always plays an integral role.

    Cell segmentation, which means determining cell boundaries in images, is an important prerequisite in modern biomedical research. However, such work generally requires experimenters to comprehensively observe all cells in a specific area for classification, counting, and analysis to obtain statistically significant conclusions [5, 6, 7]. This is tedious, time-consuming, and error-prone when conducted completely manually by pathologists. Fortunately, the increasing utilization of deep learning in

medical image processing partly solves these problems. Neural networks can extract deeper local characteristics and relationships of the cell shapes, placing them at the forefront of cell segmentation algorithm research [8]. Deep learning methods in cell segmentation are mainly based on convolutional neural networks (CNNs) [9]. CNNs enable the network to extract low-level features and advanced semantic features from multiple scales [10, 11, 12]. In recent years, Transformers [13] and Attention models [14] have also shown their potential in this area. Deep learning methods have become the trend in cell segmentation research.

For biological and medical researchers, there are already many algorithms that can help them solve cell image processing problems [15, 16, 17]. However, researchers may not have a full understanding of which algorithm is most suitable for their tasks [18, 19]. This is due to the presence of artifacts in cell images that occur during cell collection, as a result of misconfiguration of the microscope and improper operation of the researcher [19]. For bright field microscopy and fluorescence microscopy, there are three main types of common aberrations present during cellular imaging. The first type is the aberration of the microscopic imaging instrument itself, which is caused by non-ideal optical components and minor optical path adjustment errors and is mainly in the form of low-order coma, astigmatism, and trefoil error. The second type is an aberration caused by the mismatch between the refractive index of the microscopic objective lens and the biological sample, which is mainly in the form of low-order spherical aberration with amplitude proportional to the imaging depth. The third type is the aberration caused by the uneven refractive index inside the sample, where the wavefront is distorted by the highly complex biological and optical environment. The fourth type is grid distortion, which occurs primarily in large field of view (FOV) imaging. As the FOV increases, optical systems often experience geometric distortions, causing originally straight lines or grids to appear curved or warped. This type of aberration is particularly challenging in applications requiring high precision across wide areas of the sample. Both the third and fourth types are classified as mixed aberrations because they involve multiple sources of distortion that interact in complex ways. The uneven refractive index and the large FOV not only introduce distinct types of wavefront aberrations but also exacerbate each other, leading to a compounded effect that is difficult to correct with standard optical methods [20, 21]. Those aberrations can severely impact the segmentation quality and accuracy of cell images captured by bright field microscopy and fluorescence microscopy. For instance, spherical aberration can cause blurring of cell edges, making it difficult to accurately classify and count cell lineages, where precise cell boundary detection is essential for tasks such as determining cell size and morphology [22]. Astigmatism can further distort cell shapes, leading to errors in lineage assignment and tracking [23]. Coma aberration can result in the misalignment of fluorescent signals from different channels, thereby distorting the localization of nuclear markers and leading to inaccurate physiological studies, such as in the analysis of protein expression or cellular interactions. These segmentation inaccuracies can lead to significant errors in downstream analyses, such as incorrect quantification of cell populations, misinterpretation of cellular states, and unreliable results in studies requiring precise spatial resolution, ultimately compromising the validity of experimental outcomes [24].

Aberrations in bright field microscopy and fluorescence microscopy can be partly solved by delicate hardware design or software corrections. Hardware solutions, mainly entail choosing perfectly matched objectives (water or glycerol lens with certain focal length and numerical aperture, etc.), use of apochromatic lenses (chromatically corrected for multiple wavelengths) [25], or use of the wavefront detector and an adaptive lens [26]. Such strategies, however, need a comprehensive understanding of the experimental tasks and the whole optical system [27], which is sometimes overlooked. Moreover, the cost of extra hardware necessary to correct all the aberrations can be prohibitive or hard to justify when the associated image quality improvements are minor. Another choice is the deep learning method [28]. It can learn the aberration patterns using actual calibrated images or synthetic images [29]. Still, like the segmentation models, these methods also depend on the datasets, and a neural network that can help solve all kinds of aberrations is still under investigation. All in all, researchers are often hesitant to use cell instance segmentation algorithms directly for image segmentation when hardware and software methods are limited. Therefore, it is important to benchmark the robustness of cell image segmentation models under different optical aberrations and provide guidelines for their use in cell instance segmentation. The proposed guideline will inform researchers in what cases they can directly use the cell instance segmentation algorithm to segment aberration degraded images without spending more time and money to eliminate aberrations, and how to select the most appropriate cell segmentation model for different aberration degraded images. In addition, considering that cell morphologies used in bright field microscopy and fluorescence microscopy imaging tasks is not the same, and that different cell morphologies may also have an impact on the selection of segmentation models, the guideline will also give recommendations on the selection of cell segmentation models for different microscopy imaging techniques and cell morphologies. By referring to the proposed guideline for cell segmentation studies, researchers will reduce the impact of aberration degradation on the quality and accuracy of cell segmentation.

In bright field microscopy and fluorescence microscopy, image quality and aberration are usually characterized by their point spread function (PSF). Researchers can obtain the PSF image of the microscope system by using fluorescent microspheres with diameters ranging from 20 to 200 nanometers as point light sources [30]. Based on the types of aberrations displayed in the PSF images and referring to our proposed usage guidelines, researchers can reasonably apply different cell segmentation algorithms to address various aberrated cell images. However, considering that not all microscope users have undergone rigorous optical training to identify aberration types and that PSF images of different aberration types can be difficult to distinguish visually under small amplitudes, we propose a model called the Point Spread Function Image Label

Classification Model (PLCM). PLCM can quickly and accurately assist researchers in identifying the types and amplitudes of aberrations from PSF images, thus enabling better utilization of our proposed cell segmentation guidelines under various aberrations.

**Results**

**Aberration-degraded cell image analysis.** To make our proposed guideline have wider applicability, we compared common publicly available cell datasets captured by fluorescence microscopy and bright field microscopy, respectively, and selected two representative datasets from them, i.e., the DNN dataset and the LIVECell dataset, which comprise the overall dataset we used (see Supporting Information for details). In terms of cell morphology, cells in the DNN dataset have similar characteristics to the pixel features of the neighboring cells, with the same shape, and are separated from each other in the image with clear edges and less interference between cells. These properties enable cell instance segmentation algorithms to more easily distinguish and segment each cell, which are usually considered as simple cell images [31]. In contrast, cells in the LIVECell dataset are usually characterized by inconsistent morphological attributes, uniform appearance, and tightly connected cells with blurred edges, or cells overlapping each other. These factors make cell instance segmentation more difficult because the algorithms need to deal with more blurred and interfering information, which is usually considered complex cell images [32]. By categorizing the cell morphology, we can select suitable cell instance segmentation algorithms for different morphologies of the image. It is important to note that the algorithm tested is selected based on cellular morphology rather than the device used to capture the cells (see Supporting Information for details).

After specifying the overall dataset, we introduced various types and degrees of aberrations to the images, resulting in multiple forms of simulated aberration-degraded image test sets. The detailed process is shown in Figure 1. Given the optical parameters of the microscope and the amplitudes of Zernike polynomials, various PSFs were generated to represent different aberrations. These PSFs were then convolved with the unaberrated cell images, resulting in degraded images with different aberrations.

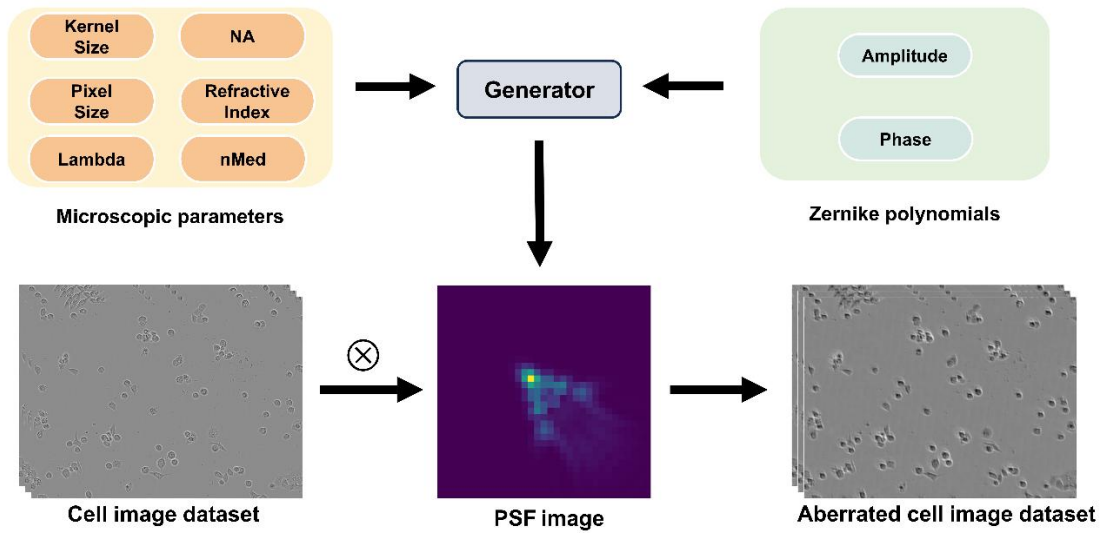

Figure 1: The process diagram for generating an aberrated cell image dataset by inputting microscopic parameters and Zernike polynomials information.

From Figure 2a, we observed that as the aberration amplitudes increase, distinct differences emerge among PSF images corresponding to different aberration types. Astigmatism results in elliptical PSF images, Coma forms "teardrop" shapes, Spherical aberration diffuse images with ring-like structures, and Trefoil in PSF images is shown as three main "lobes". These differences are also reflected in cell images, with each aberration type exhibiting distinct degradation patterns. Figure 2b further analyzes PSF images with mixed aberrations, showing that as the number of orders increases, PSF kernel size and differences between PSF images also increase. However, differences among cell images with mixed aberrations are not significant, posing challenges for visual distinction.

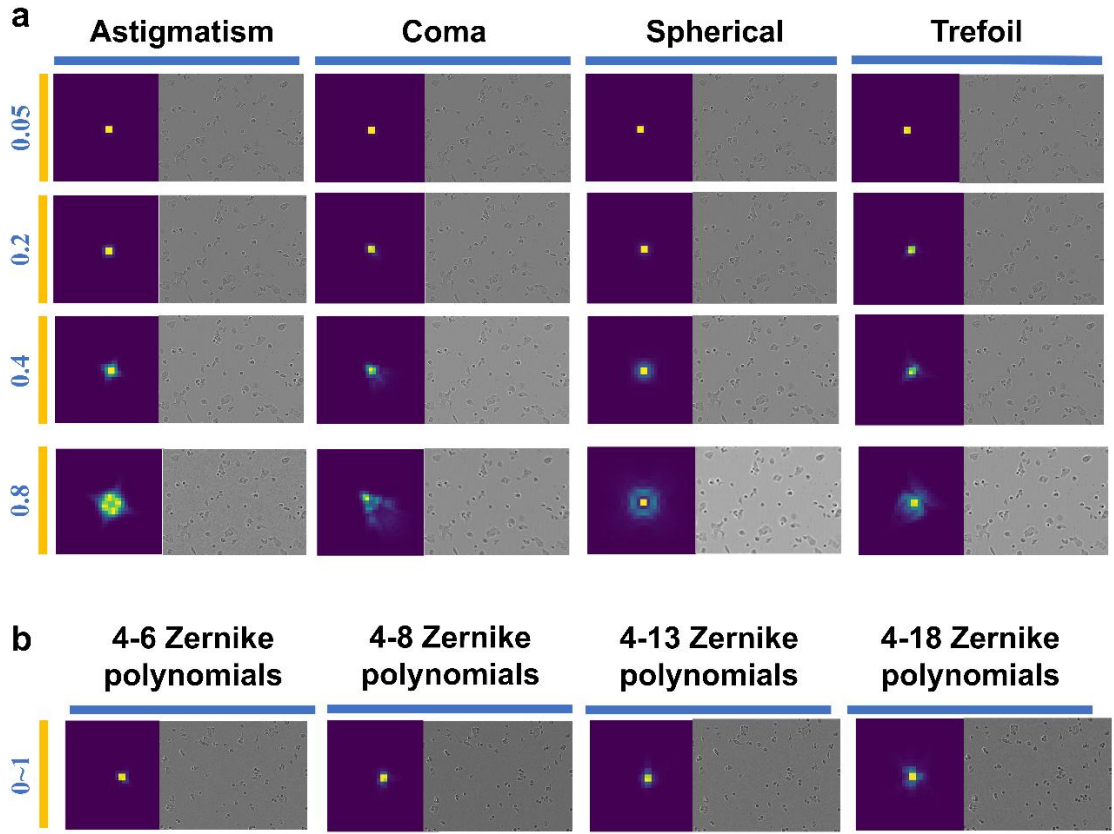

Figure 2: Aberration-degraded cell image analysis. a-b, The PSF images with various aberrations and their corresponding BT474 cell images from the LIVECell dataset. (a) Single aberrations. (b) Mixed aberrations.

In Figure 3, we conducted a numerical analysis using PSNR, SSIM, and correlation coefficient as similarity metrics to validate the experimental results presented in Figure 2 and to evaluate whether different cell morphologies would yield similar results. Figure 3 demonstrates that as the severity of aberrations increases, the similarity between these cell images gradually decreases, and the differences become more pronounced. In the evaluation of single aberrated cell images, we observed a significant trend: when aberrations are small, all three metrics show a sharp decline, but when the aberration amplitude exceeds 0.4, these metrics tend to stabilize. The evaluation results of mixed aberrated cell images indicate that changing the order of Zernike polynomials has a certain impact on the similarity metrics between cell images, but these effects are not significant. Additionally, these results are consistent across datasets with different cell morphologies, suggesting that our evaluation method is transferable across various cell datasets.

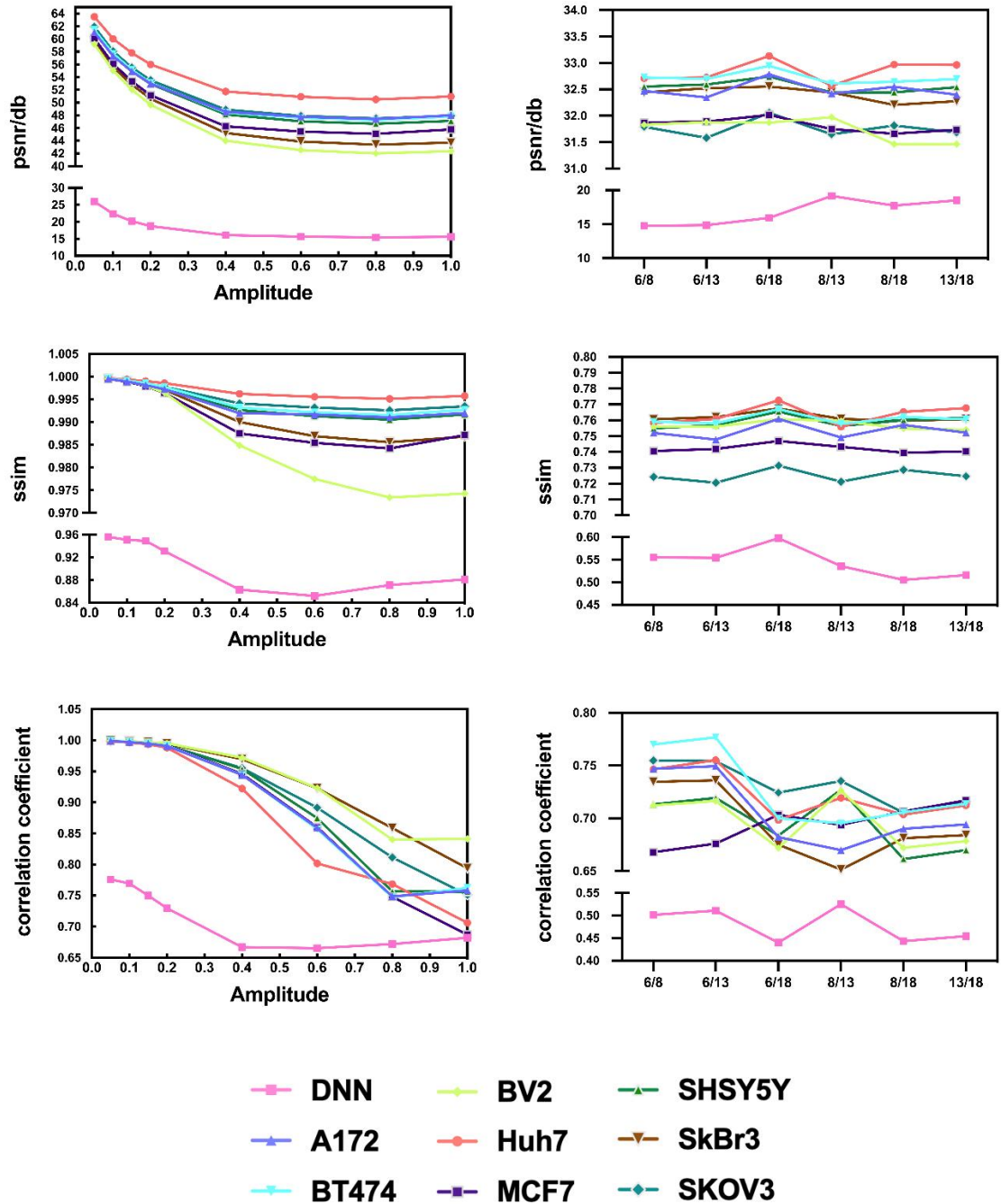

Figure 3: The trends of PSNR, SSIM, and correlation coefficient values with respect to amplitude variations among different types of single aberration cell images. And the changes in PSNR, SSIM, and correlation coefficient values across various orders of Zernike aberration cell images.

    To further validate and investigate the results, we simulated the optical transfer function (OTF) under various aberrations using the optical parameters of the microscope system and Zernike polynomials (see Materials and methods for detail). By taking the magnitude of the OTF, we obtained the modulation transfer function (MTF), which measures the contrast transmission capability of the optical system at different spatial frequencies. The MTF is commonly used for analyzing and evaluating the imaging quality of the system. As an example, we have illustrated the MTF of the optical system under different aberrations using the DNN dataset in Figure 4.

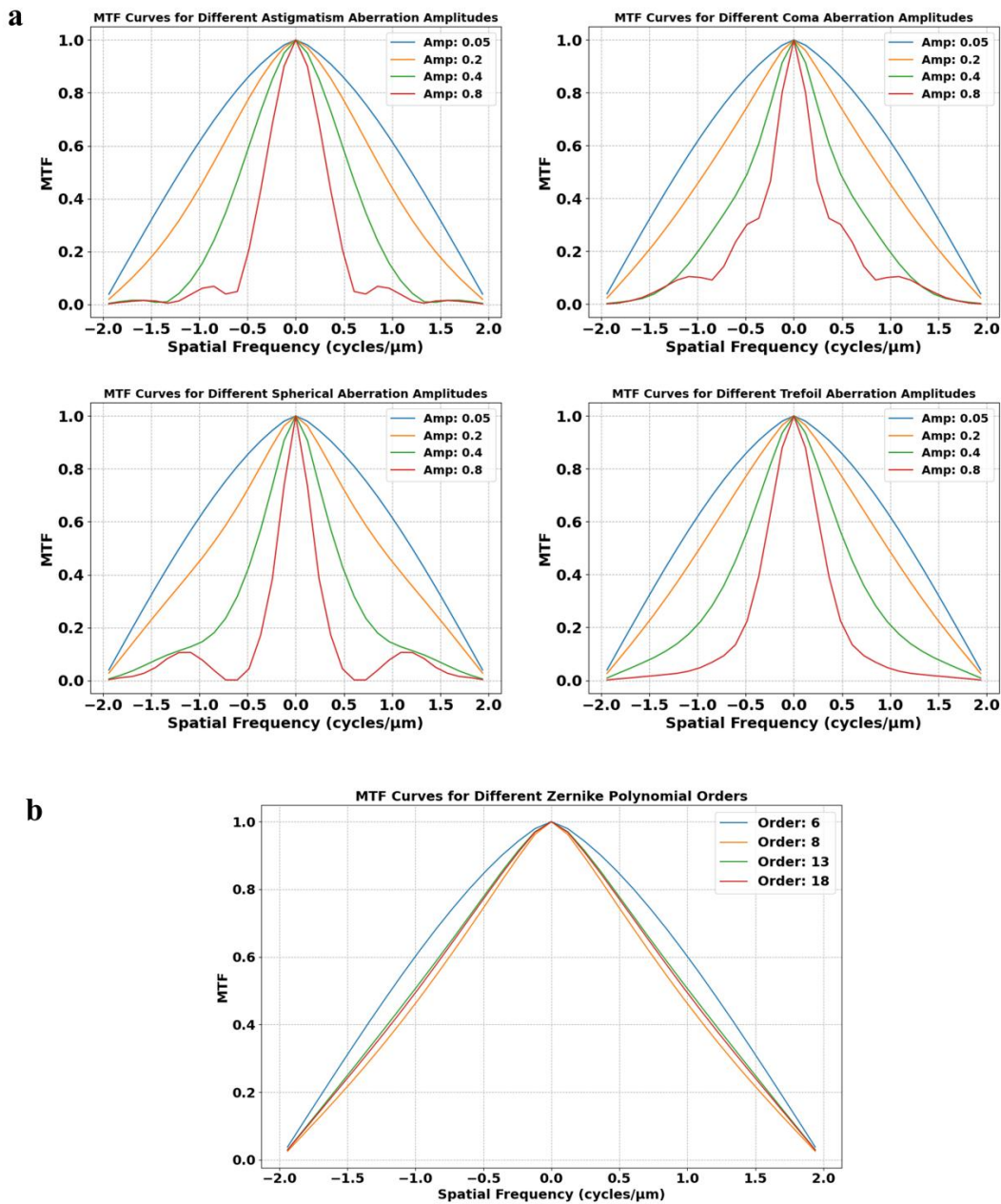

Figure 4: Modulation Transfer Function (MTF) curves illustrating the effects of varying aberration magnitudes and Zernike polynomial orders. (a) MTF curves for different aberration magnitudes. (b) MTF curves for different orders of Zernike polynomials under mixed aberration conditions.

From Figure 4a, we can observe that when the aberration magnitude is small, the MTF curve shows a gradual decrease in the spatial frequency response, but the high-frequency part (which represents the details in the image) is still well preserved, and the difference in the MTF curves between different aberration species is not significant. This means that the PSF is still effective in maintaining high spatial frequencies, resulting in a very limited quality degradation of the aberration image, and there is no significant difference between the degraded images of different aberration species. As the aberration magnitude increases, there is a rapid degradation of the spatial frequency response, especially in the high spatial frequency part. This leads to a significant reduction in the quality of image details, which triggers a more pronounced image degradation effect. At this point, the differences in the MTF curves at different aberrations become more pronounced, corresponding to the increasing effect of aberration on image quality. This implies that the PSF is more diffuse and affects a wider frequency range. Figure 4b shows that the MTF curves do not show significant differences between different orders of Zernike polynomials in the mixed aberration case using Gaussian sampling to simulate a realistic scene. The MTFs are relatively consistent, and the differences in spatial frequency response are small, which is consistent with the observed image degradation differences between the different orders of Zernike polynomials. This corresponds to the observation that there is little difference in image degradation at different Zernike polynomial orders.

**Network heads and Network backbones robustness evaluation.** Cell images with different or similar pixel features to neighboring cells are described as simple cell images, and segmentation tasks for such images are very common in cell segmentation. When tackling instance segmentation tasks on simple cellular images, researchers often opt for various combinations of network heads and network backbones. They then proceed to fine-tune the parameters on their cell datasets to establish these models as state-of-the-art within their respective datasets. Therefore, testing the robustness of bounding box recognition and the suitability of a selected backbone for a given network head is necessary. In this section, we take Mask R-CNN as the framework and use the convenience brought by its modular structure to replace different network heads for bounding box recognition and backbone to test its prediction performance on different aberration-degraded DNN datasets. We report average precision (AP) as our metric for instance segmentation [33].

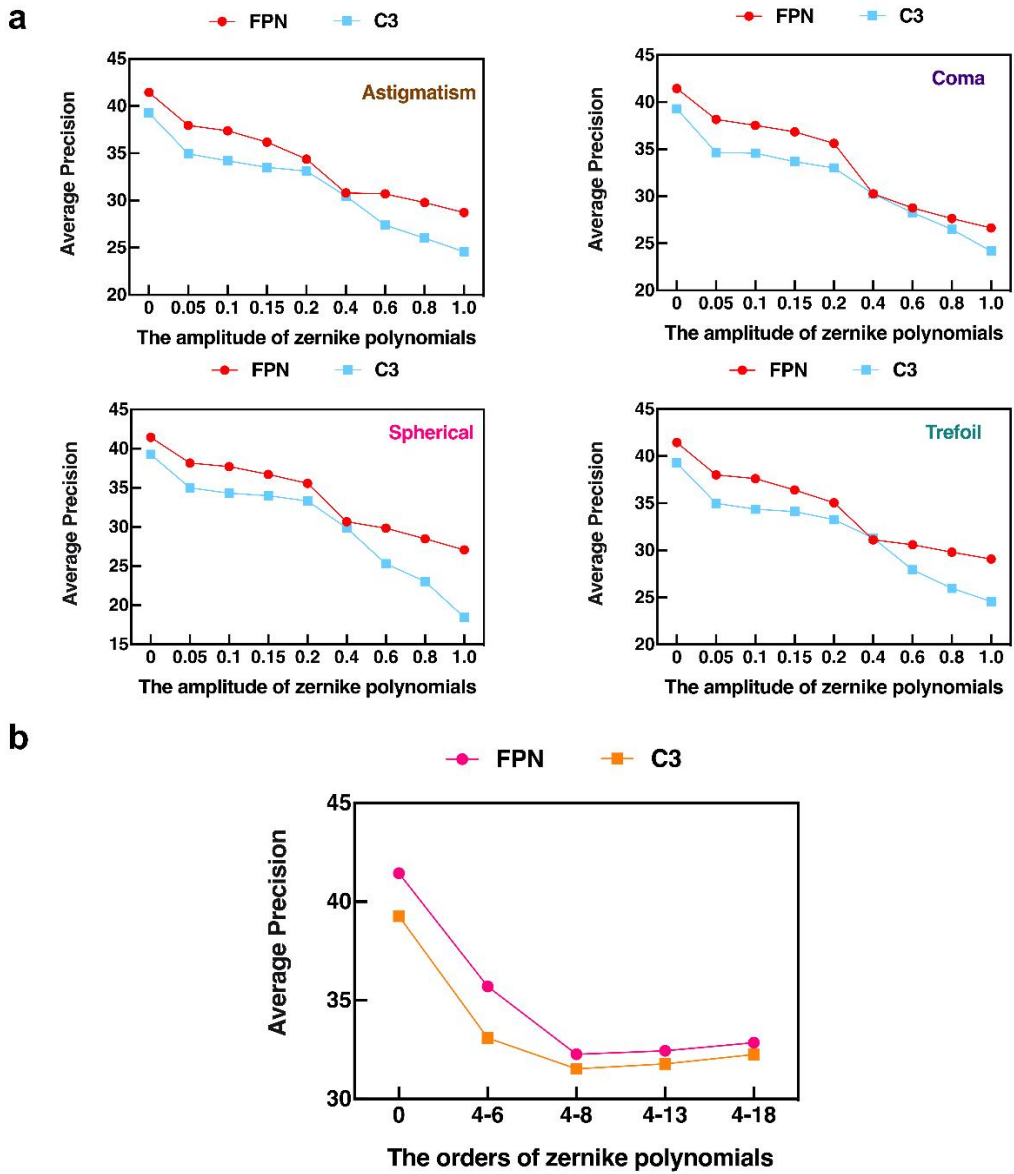

Figure 5: The test results of AP values for C3 and FPN on aberration-degraded images of the DNN dataset. (a) Single aberrations. (b) Mixed aberration

We employed ResNet50 as the backbone and varied network heads (FPN [34] and

C3 [35]) for bounding box recognition in constructing Mask R-CNN models. FPN, the Feature Pyramid Network, integrates features across multiple scales via a top-down pathway with lateral connections, while C3 extracts features from the final convolutional layer of the 3rd stage. Training details are outlined in the Materials and Methods section. Testing on the DNN test set with different aberration degradations revealed the superior performance of FPN over C3, as shown in Supporting Information Table S5 and Figure 5a. FPN's integration of bottom-up feature extraction, top-down feature propagation, and lateral connections facilitated better utilization of hierarchical information, outperforming C3 across various aberration types and amplitudes. Furthermore, we observed that both C3 and FPN exhibit higher similarity in AP values on Astigmatism, Coma, Spherical, and Trefoil aberrations. This indicates that altering aberration types while maintaining amplitude does not significantly affect AP values. The performance of FPN and C3 on mixed aberration degraded images is depicted in Figure 5b, where FPN surpasses C3 under the influence of various mixed aberrations. FPN achieves an AP value exceeding 35 in the case of 4-6 orders of Zernike polynomials, while accuracy declines in other scenarios. It is noteworthy that as the orders of Zernike polynomials increase from 4-8 to 4-18, both FPN and C3 exhibit an upward trend in predicted values, which seems to contradict the relatively insignificant differences observed among cell images with different mixed aberrations, as depicted in Figure 2b. However, we observed that the difference in AP values does not exceed 0.7 in the cases of 4-8, 4-13, and 4-18 orders of Zernike polynomials. Hence, we can infer that FPN and C3 yield nearly identical prediction results on degraded images with the latter three orders of Zernike polynomials.

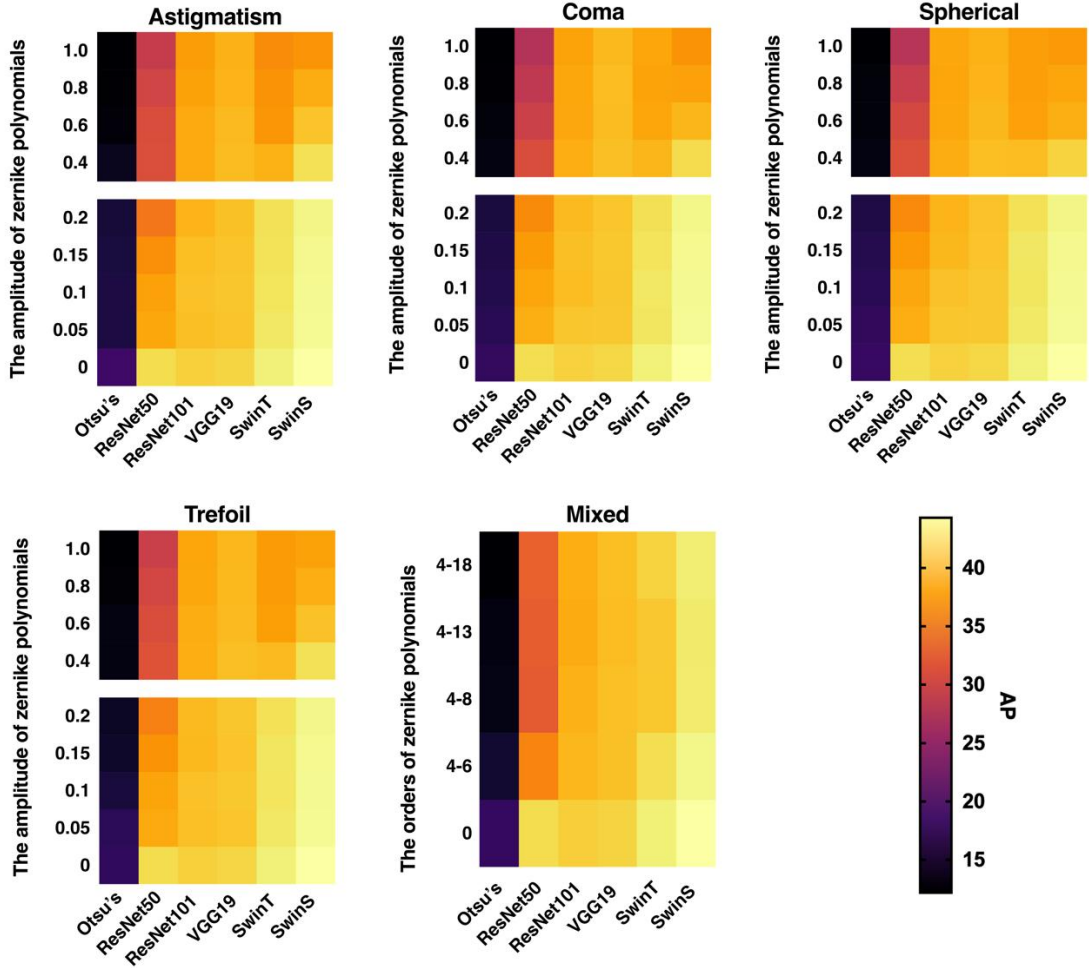

Figure 6: The segmentation performance of different backbones under single and mixed aberrations, using the DNN dataset.

We selected FPN as the network head for bounding box recognition and tested the performance of various backbone networks (ResNet50 [36], ResNet101, VGG19 [37], SwinT [38], SwinS). Additionally, we included the Otsu threshold method (Otsu)[39], a traditional cell segmentation approach, as a baseline for comparison. AP results for different backbones are presented in Supporting Information Table S6. In Figure 6, we observed that Swin transformer architecture, particularly SwinS (96M), exhibited the highest predictive performance when the amplitude ranges from 0 to 2. As amplitude increased, Otsu, ResNet, and Swin transformer architectures experienced a decline in predictive performance, with VGG19 displaying greater stability.

Figure 6 also illustrates the performance of different backbones under mixed aberrations. While altering the orders of Zernike polynomials minimally affected backbone performance, aligning with our earlier conclusion on the insignificance of differences between cell images with mixed aberrations, SwinS consistently outperformed other backbones in mixed aberrated image evaluation.

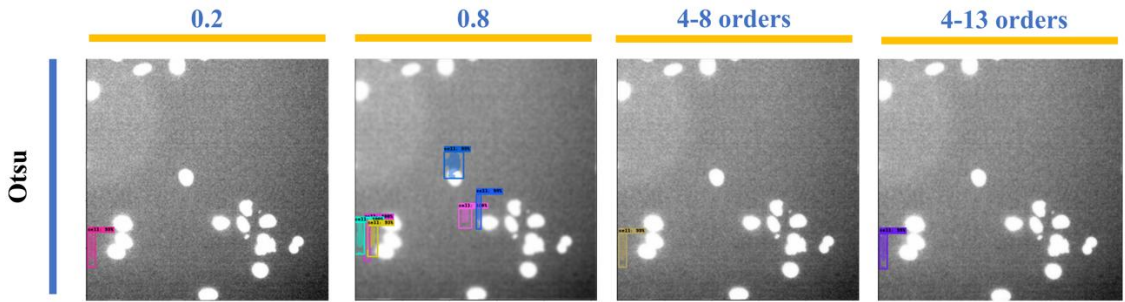

Figure 7: The predictive performance of the Otsu algorithm using aberrated DNN image as an example.

Figure 7 shows the performance of the traditional Otsu algorithm for cell segmentation under different aberrations. The results clearly show that the Otsu algorithm can not achieve effective cell segmentation in either single aberrated cell image or mixed aberrated cell image. Under different amplitudes and different orders, it is difficult for the algorithm to distinguish the cell boundary from the background. These findings highlight the shortcomings of the algorithm when dealing with aberrated imaging conditions, highlighting the need for more powerful deep-learning segmentation techniques in such situations.

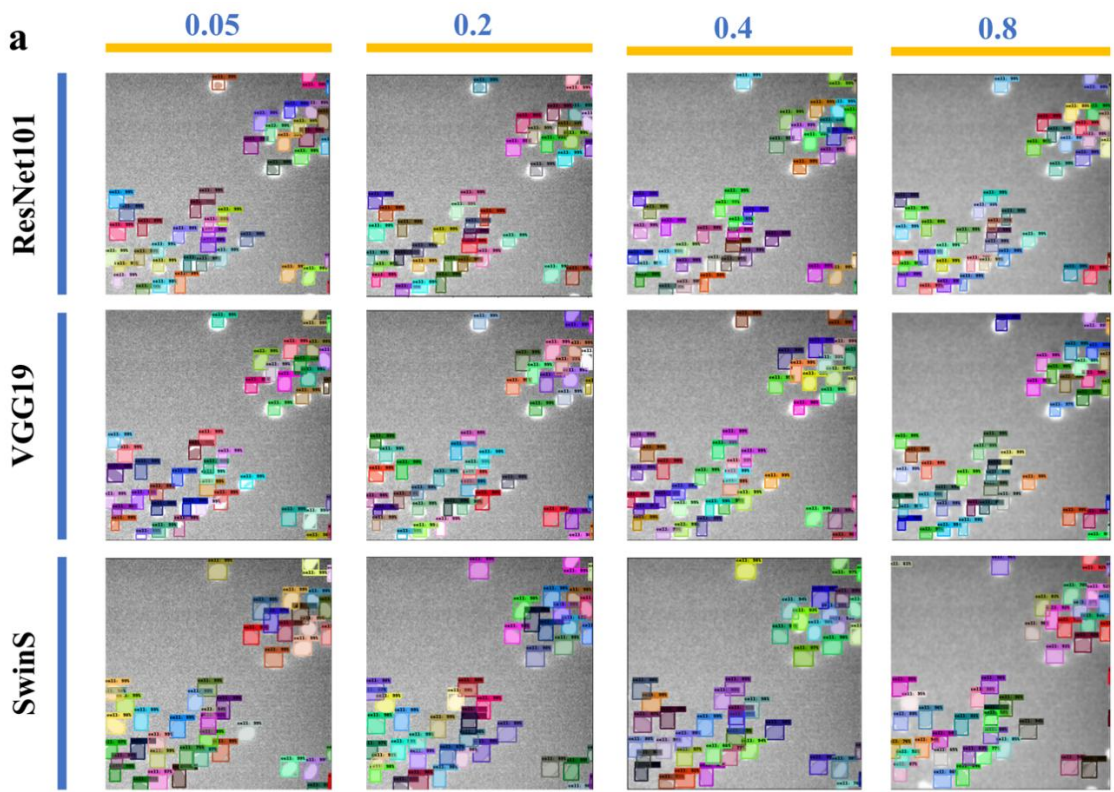

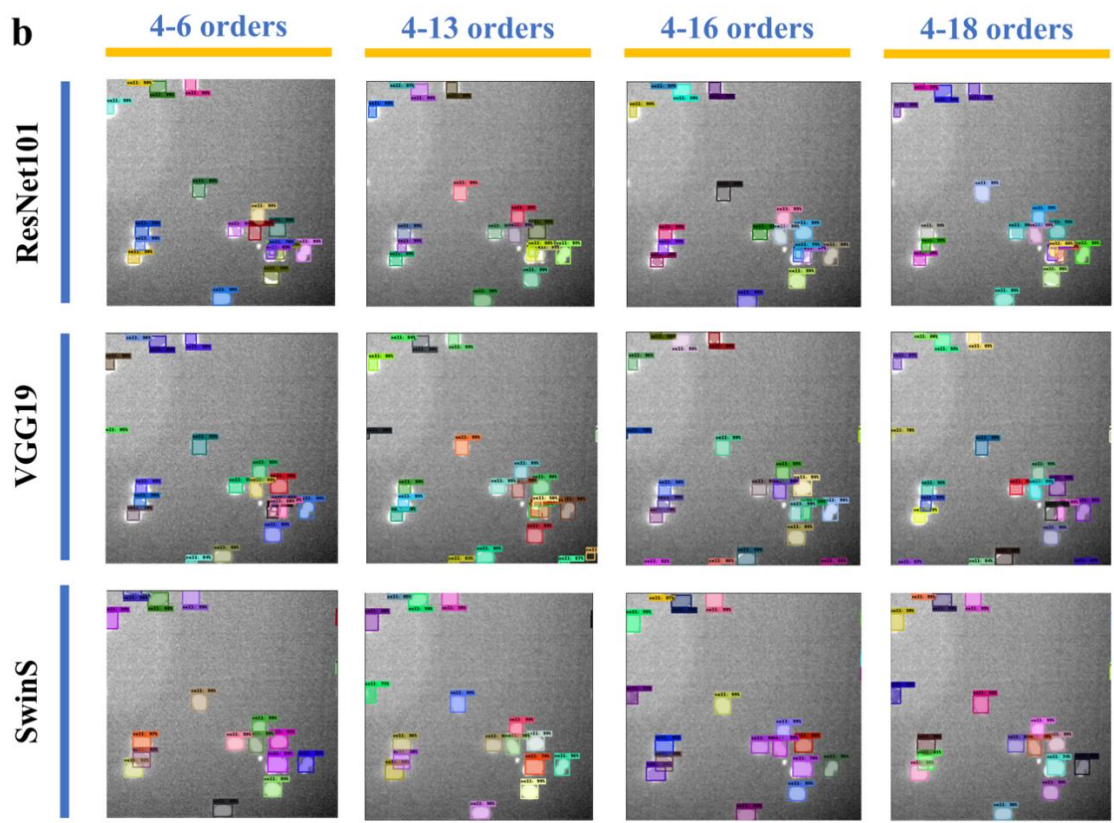

Figure 8: The predictive performance of ResNet101, VGG19, and SwinS. (a) Using the single aberrated (Astigmatism) DNN image as an example. (b) Using the mixed aberrated DNN image as an example.

In Figure 8a, we observed that SwinS consistently delineated cell boundaries accurately, while ResNet101 and VGG19 occasionally miss segmented boundaries, particularly as amplitude increased. However, SwinS also faced challenges with increased degradation, with a notable decrease in cell detection rates. Consequently, VGG19 outperformed SwinS in mean average precision under extreme degradation conditions. Figure 8b, exemplified by ResNet101, VGG19, and SwinS, showcases prediction performance on mixed aberration degraded images. SwinS demonstrated accurate and complete cell segmentation, whereas ResNet101 and VGG19 exhibited incomplete segmentation, edge miss segmentation, and cell overlap. Therefore, for straightforward cell images with mixed aberrations, SwinS is recommended as the backbone for segmentation, eliminating the need for additional aberration correction efforts.

Therefore, for both single and mixed aberration cell images, we strongly recommend using the FPN+Swin Transformer architecture to build a deep learning network for cell segmentation tasks. Furthermore, when degradation is more severe, optical equipment or post-processing methods should be considered to eliminate aberrations. Furthermore, due to the degradation in image quality caused by aberrations, the segmentation task becomes more complex, necessitating more powerful feature extraction capabilities to ensure segmentation accuracy. We recommend that users should have at least a GPU with 12GB or more VRAM, such as the NVIDIA RTX 3080 or higher models, or utilize high-performance computing instances available on cloud platforms to effectively run these models. In the Supporting Information, we discuss the specific computational costs of the aforementioned network heads and network backbones and provide constructive suggestions to help minimize these costs.

**Cellpose2.0 Toolbox robustness evaluation.** Cell instance segmentation algorithms aim to partition cell images into individual cell instances. This can be challenging due to various factors such as inconsistent morphological properties, homogeneous appearances, and tightly crowded cells. These complexities make traditional segmentation algorithms less effective and often require manual intervention for accurate results. Deep learning toolboxes like Cellpose2.0 [40] offer significant advantages over traditional segmentation algorithms for handling complex cell images, including improved accuracy, scalability, adaptability, and user-friendliness. Therefore, it is essential to conduct robustness testing of the segmentation performance of Cellpose2.0, the most representative and widely used deep learning toolbox for complex cell segmentation, under various aberrations and for different cell morphologies.

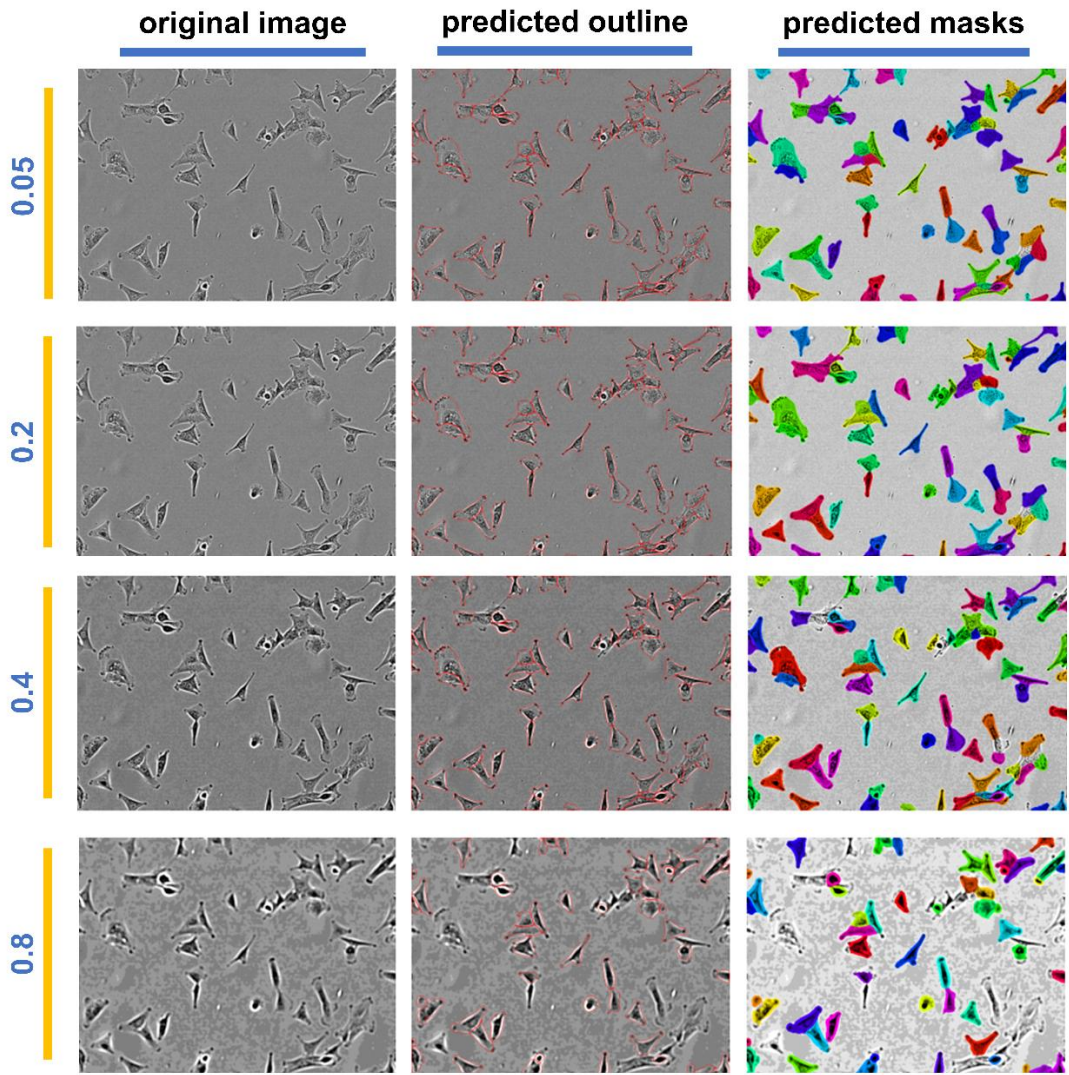

Figure 9: The performance of Cellpose2.0 under varying amplitudes (AP50, A172 cell, Astigmatism aberration).

We utilized the live-cell model weights of Cellpose2.0 for testing and selected the anchor-free model from the LIVECell dataset as the baseline, assessing instance segmentation using AP50 and AP75 metrics. Supporting Information Table S7 presents predictive values for single-class aberrated images, indicating that changing the aberration type under the same amplitude insignificantly affects prediction results, across eight different morphological cell images. In Figure 9, we observed that Cellpose2.0 exhibited excellent segmentation performance with amplitudes up to 0.2, accurately predicting cell contours and segmented regions. However, increased amplitude significantly impacted segmentation performance, leading to incorrect predictions and segmentation failures under extreme degradation. Consequently, for complex cell image segmentation, we recommend using Cellpose2.0 directly without aberration elimination, provided that the aberration amplitude is minor.

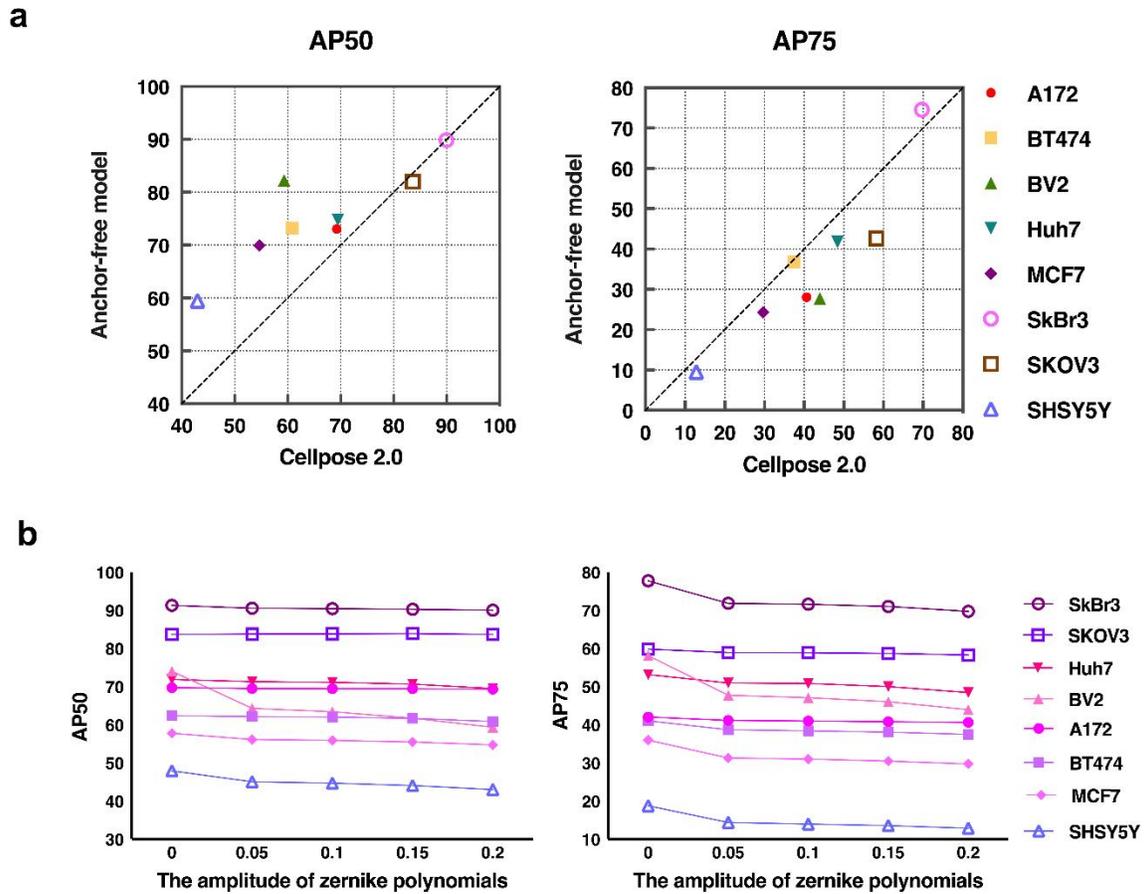

Figure 10: Cellpose2.0 toolbox robustness evaluation of single aberrations, taking Astigmatism aberration as an example. (a) AP50 and AP75 value (amplitude: 0.2) (b) The trend of changes in AP50 and AP75 values as the amplitude varies.

In experiments using AP50, Cellpose2.0 showed similar performance to the anchor-free model in SkBr3 and SKOV3 cell types but trailed behind in the other six types. Taking the BV2 cell with the largest difference for example, although the Cellpose2.0 AP50 value lagged behind the anchor-free model by approximately 23, it still reached 59, indicating 3that the Cellpose2.0 AP50 value is still within an acceptable range. With AP75, Cellpose2.0 outperformed the anchor-free model in all types except SkBr3, suggesting its suitability for complex cell image segmentation, particularly when using AP75. The results are shown in Figure 10a. Further analysis in Figure 10b reveals the performance of Cellpose2.0 variations across different cell types under Astigmatism aberration, where SkBr3 and SKOV3 cells exhibit the best results and SHSY5Y cells show the poorest performance.

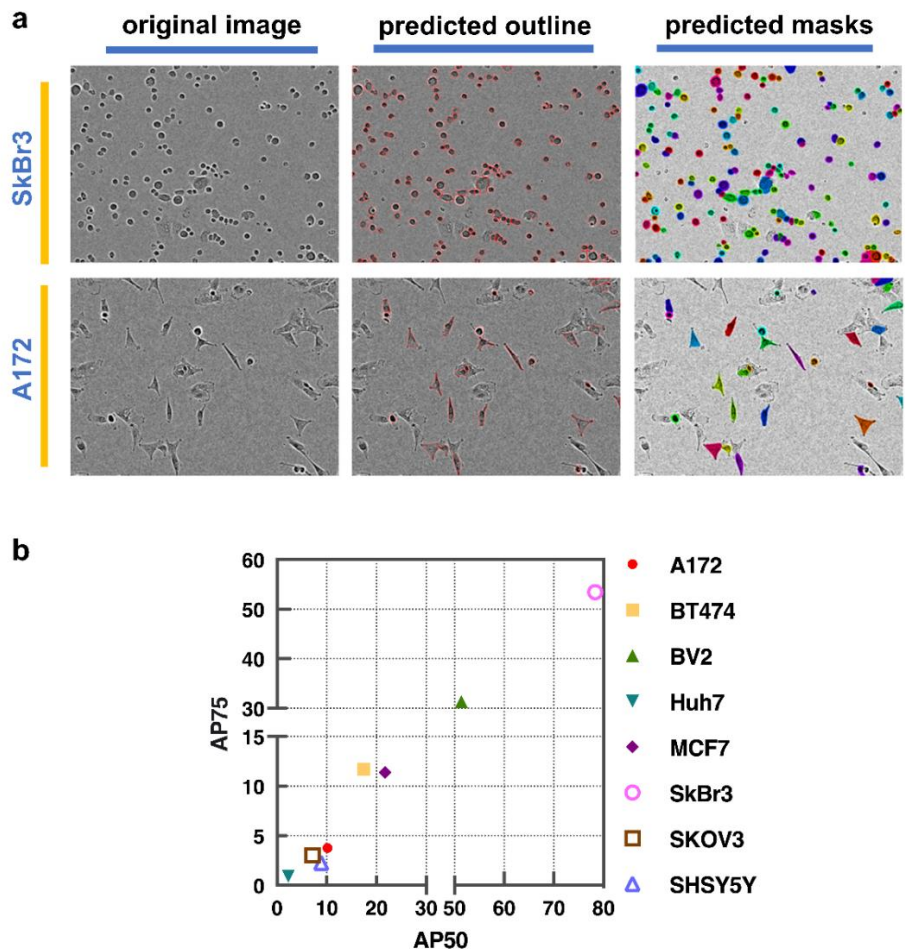

Figure 11: Cellpose2.0 toolbox robustness evaluation of mixed aberrations, taking 4-8 orders aberration as an example. (a) The segmentation performance of Cellpose2.0 on SkBr3 and A172 cell images. (b) AP50 and AP75 values across the 8 classes of cell images.

Supporting Information Table S8 presents predictive values for mixed aberrated images. Across all 8 classes of cell images, Cellpose2.0 performs consistently under mixed aberrations with different orders of Zernike polynomials, with minimal differences in AP50 (not exceeding 10) and AP75 (not exceeding 8). However, significant performance discrepancies exist across different cell types, with Cellpose2.0 demonstrating good segmentation performance only on BV2 and SkBr3 cells. In Figure 11a, Cellpose2.0 effectively identifies and segments most cells in SkBr3 images, but it struggles with A172 images, segmenting only a few cells accurately and rendering it unusable in such cases. Figure 11b presents that Cellpose2.0 achieves satisfactory results only for BV2 and SkBr3 cells, with AP50 values below 30 and AP75 values below 15 for other cell types. The usability of Cellpose2.0 is limited to BV2 and SkBr3 cell types under mixed aberrations.

**Point spread function image label classification model (PLCM).** This model learns image features from PSF images and uses them to predict three labels: category, aberration types, and amplitude. The category includes single aberration and mixed aberration.

The proposed PLCM model can effectively learn aberration features from PSF images in an end-to-end manner. The model consists of two parts: (1) image feature learning component and (2) three classification heads for different labels. The first part is composed of a CNN network, which can automatically learn features from PSF images. The second part consists of three classifiers, each with its cross-entropy loss, which can evaluate the prediction accuracy of the model for different types of labels. By replacing the CNN model in the feature extraction module with popular feature learning models such as MobileNet [41], EfficientNet [42], SqueezeNet [43], and DenseNet [44], and conducting training and testing experiments on the PSF image dataset, we aim to find the most suitable model for PSF image feature extraction tasks (see Materials and Methods for details for training and testing experiments).

**Table 1:** Performance comparison of PLCM models composed of different CNN

networks.

| Model | Category | | Aberration Type | | Amplitude | | Parameters |
|---|---|---|---|---|---|---|---|
| | Accuracy | Precision | Accuracy | Precision | Accuracy | Precision | |
| MobileNet | 63.28% | 86.27% | 55.84% | 83.93% | **58.77%** | **81.90%** | 2227715 |
| EfficientNet | 61.74% | 81.50% | 55.22% | 80.28% | 57.81% | 80.04% | 4011391 |
| DenseNet | **63.96%** | **89.98%** | **58.47%** | **85.36%** | 58.23% | 81.75% | 6956931 |
| SqueezeNet | 48.40% | 72.87% | 41.63% | 69.57% | 47.28% | 71.59% | 736963 |

The results presented in Table 1 demonstrate that DenseNet achieves the highest prediction accuracy and precision in terms of Category and Aberration Type, while MobileNet demonstrates excellent prediction performance in Amplitude. SqueezeNet, on the other hand, has the least training parameters and computational overhead, resulting in the poorest prediction performance. It should be noted that the difference in prediction performance between MobileNet, EfficientNet, and DenseNet is not significant. However, MobileNet has the least training parameters among the three. Additionally, as MobileNet is a lightweight network specifically designed for mobile and embedded devices, its core feature is the use of depthwise separable convolution. This convolution method significantly reduces computation and the number of parameters compared to traditional convolution, making it more friendly to researchers with limited computational resources. In summary, we choose MobileNet as the feature extraction module, and the architecture of PLCM is illustrated in Materials and Methods.

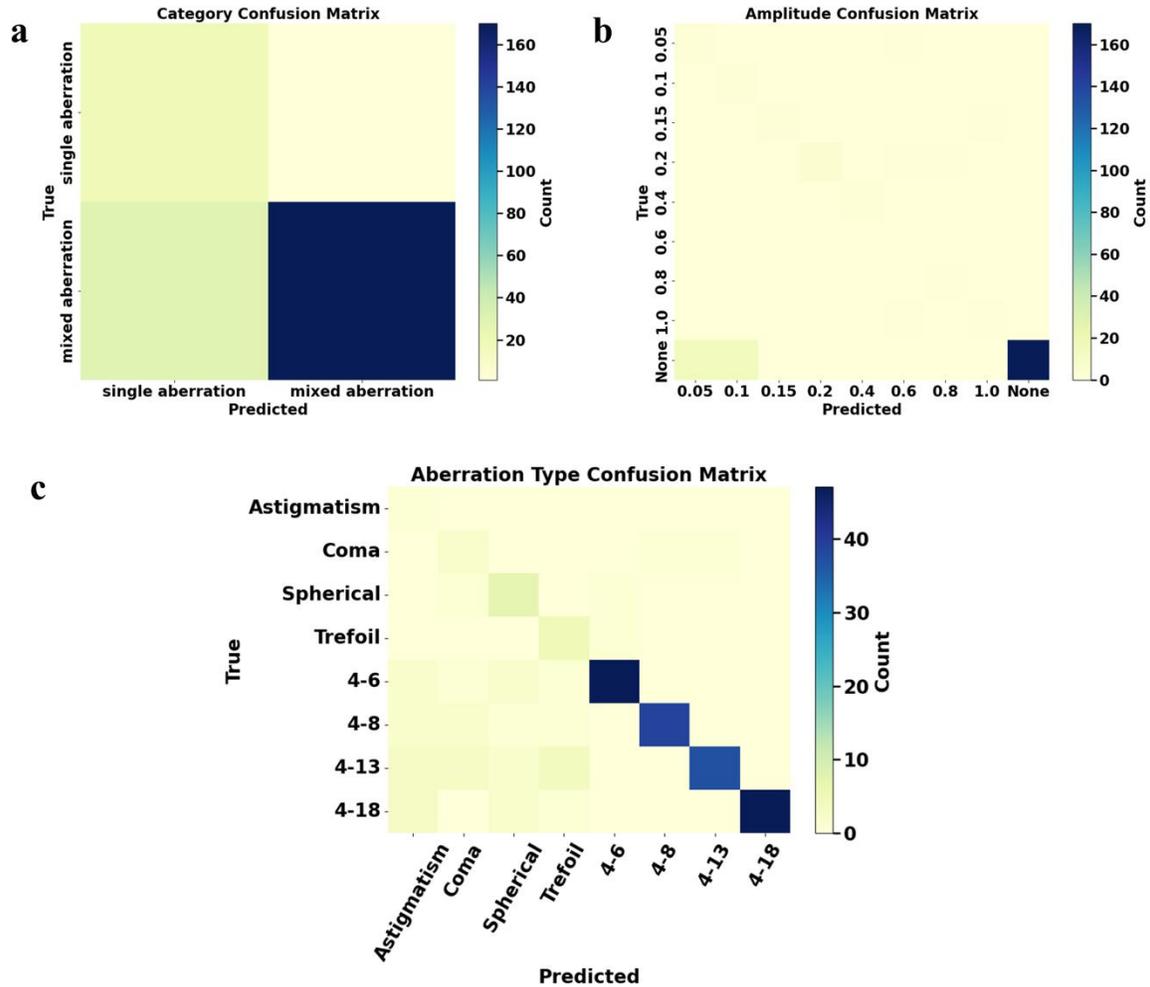

Figure 12: Confusion matrices for the PLCM classifier.

The confusion matrices shown in Figure 12 provide a comprehensive evaluation of the classification performance of the PLCM classifier across three labels: Category, Aberration Type, and Amplitude. Figure 12b illustrates that the amplitude prediction accuracy is notably high, with PLCM effectively distinguishing between various amplitude levels. Figure 12c reveals that when the amplitude is between 0.1 and 0.2, there is significant confusion among the four types of single aberrations—Astigmatism, Coma, Spherical, and Trefoil—indicating challenges in differentiating these aberrations under intermediate amplitude conditions. Additionally, Figure 12c shows that at amplitudes below 0.1, single aberration types frequently overlap with the mixed aberration category, particularly the 4-6 mixed aberration type, leading to increased misclassification between single and mixed aberrations. This phenomenon highlights the difficulty PLCM experiences in distinguishing between these categories at low amplitudes. Meanwhile, Figure 12 demonstrates that outside these specified ranges, PLCM maintains high accuracy, effectively distinguishing between categories, aberration types, and amplitudes. Overall, these matrices underscore the strengths of PLCM in specific conditions and reveal areas for improvement, particularly in handling low amplitude cases and distinguishing between similar aberration types.

**Discussion**

**Aberrated cell images.** Figure 2 suggests that differences between cell images with various aberrations are insignificant when aberration amplitudes are minor. However, as amplitudes increase, differences become more pronounced. This is attributed to the effective kernel size in the PSF image during convolution, representing the portion truly affecting image degradation (central region in Figure 2a). When amplitudes are below 0.2, effective kernel sizes remain largely unchanged, resulting in similar outcomes across aberrations. Yet, as amplitudes rise, effective kernel sizes diversify, leading to varied degradation effects. This conclusion is also validated in the experiment with mixed aberrations shown in Figure 2b. Although the orders increase continuously, the Zernike polynomial orders at extreme amplitudes are relatively small due to the use of Gaussian sampling to simulate real-world scenarios. Therefore, there are no significant changes in the size and morphology of the effective kernel parts of the PSF image, leading to minor differences between cell images degraded with mixed aberrations of different orders of Zernike polynomials. The MTF curve in Figure 4 also verifies the above view from the perspective of quantitative analysis. When the effective kernel is small, the difference in spatial frequency response is small. When the effective kernel increases with the amplitude, the spatial frequency response decreases rapidly, especially in the high spatial frequency part, at this time, the degradation effects of different types become more obvious.

**Network heads and network backbones.** Both C3 and FPN demonstrate comparable performance across single-class and mixed aberrated images, owing to their focus on multi-scale feature extraction, cascaded strategy for accuracy enhancement, and utilization of data augmentation techniques during training. FPN's multi-scale feature pyramid enhances semantic information capture from images at varying scales, contributing to its superior performance. Network backbones such as ResNet, VGG, and Swin Transformer architectures exhibit strong feature extraction capabilities, resulting in similar segmentation performance across aberrated images. The Swin Transformer, with its self-attention mechanism and multi-scale feature fusion, excels in cell image segmentation by capturing global cell correlation and comprehensively understanding image structures. Particularly, SwinS achieves the best segmentation performance in aberration-degraded cell images due to its larger training parameters. Therefore, a combination of FPN and SwinS is recommended for segmentation tasks on aberration-degraded simple cell images.

**Cellpose2.0 Toolbox.** Cellpose2.0 demonstrates satisfactory performance in single-class aberrated image segmentation tasks on complex cell images due to its dynamic merging and segmentation strategy. This strategy handles overlaps and contacts between cells better, resulting in more accurate segmentation results. Additionally, Cellpose2.0 can handle multi-channel cell images and be applied to different cell types, making it more versatile, as demonstrated in experiments on single-class aberrated images of 8 tested cell types. In more complex and challenging mixed aberrated image segmentation tasks, Cellpose2.0 exhibits significantly different segmentation performance across different cell morphologies. When dealing with round-shaped cells (such as BV2 and SkBr3 cells), Cellpose2.0 shows better segmentation results. This is attributed to its deep learning model's ability to adaptively adjust segmentation size based on cell size and shape, thereby accommodating the morphological characteristics of round cells better.

Additionally, the self-attention mechanism of Cellpose2.0 allows the network to capture spatial relationships globally, aiding in better understanding the correlations between round cells.

**PLCM method.** The reasons for PLCM to choose MobileNet as the feature extraction module mainly include the following points. 1) MobileNet, as a feature extractor, achieves a good balance between efficiency and performance. Its depthwise separable convolution structure effectively reduces the number of parameters and computations, thereby speeding up the training and inference speed of the model without significantly reducing accuracy. 2) Although EfficientNet and DenseNet perform similarly to MobileNet in terms of accuracy, their complex compound scaling strategies and higher computational requirements make them less practical than MobileNet in resource-constrained environments. 3) Compared to SqueezeNet, despite SqueezeNet also aiming to reduce the number of parameters, its compression strategy may not perform as stably as MobileNet in some complex tasks, especially in multi-task learning. In summary, the PLCM constructed with MobileNet as the feature extraction network demonstrates good performance and relatively low computational overhead in PSF image multi-label classification tasks.

Analysis of the PLCM confusion matrices highlights two key issues. First, at low amplitudes, PLCM frequently misclassifies aberration types because the features of single aberrations and the 4-6 orders of mixed aberrations are very similar. This similarity leads to increased misclassification. Second, although misclassifications at low amplitudes do not affect the usability of the guidelines—since the robustness of the model remains consistent across low amplitude single aberrations and 4-6 orders of mixed aberrations—future work should address this issue to prevent potential confusion for users.

**Conclusion**

This paper reports on evaluating the robustness of cell instance segmentation models under various optical aberrations in fluorescence and bright field microscopy images. When using fluorescence microscopy or bright field microscopy, if the cells you are capturing exhibit pixel features similar to neighboring cells, we recommend establishing your segmentation model based on a combination of FPN and SwinS architectures and fine-tuning the model on your dataset to achieve effective cell image segmentation in the presence of such aberrations. In this scenario, the specific type of aberration introduced may not need to be explicitly considered, provided that if single aberrations are present in the images, they should be within a minor magnitude amplitude (i.e., amplitude ≤ 0.2). For cells with inconsistent morphological characteristics, uniform appearance, and tightly enclosed features, when conducting segmentation tasks on such aberrated cell images, it is advisable to use the Cellpose2.0 Toolbox directly if single aberrations are introduced, under the condition of minor magnitude amplitude. If mixed aberrations are introduced, ensure that the cell images you are capturing have pixel characteristics similar to BV2 and SkBr3 cells before using the Cellpose2.0 Toolbox directly. We also proposed a Point Spread Function Image Label Classification Model (PLCM) that utilizes MobileNet as the feature extraction module, achieving recognition of PSF image labels, including Category, Aberration Type, and Amplitude, with relatively low computational overhead. This enhances the applicability of our proposed guidelines, making them more accessible to researchers who may not have access to systematic optical training. As a result, they can better utilize our guidelines.

This study also has several limitations and areas for future improvement. While the study focused on four main types of single aberrations—Astigmatism, Coma, Spherical, and Trefoil—future work should explore other aberration types such as Defocus. The research also tested only mainstream network heads, backbones, and cell segmentation toolboxes, indicating that broader testing of various models and toolboxes is needed. Furthermore, the PLCM model shows confusion among aberration types with similar features at low amplitudes, highlighting the need for more advanced classification models as deep learning technology progresses. Overall, future research should aim to develop more adaptive, accurate, and usable guidelines for cell segmentation models under aberrated images to assist researchers in the biological field with deep learning techniques for cell image analysis.

**Materials and methods**

**Aberration Simulation.** According to previous research on simulating aberration-degraded images [45], PSF can be generated by Zernike polynomial equation simulation, and this generation method is based on the optical parameters of the microscope. The simulated aberration-degraded image dataset is useful for the training, validation, and testing of deep learning models. Therefore, the testing data for cell segmentation models compared in this study is generated by the above simulation process based on DNN and LIVECell datasets. Specifically, the optical imaging model for simulating aberration degradation image generation can be expressed as:

$$I_{\text{synth}} = S * PSF_{\text{zernike}}$$

where $I_{\text{synth}}$ refers to simulated aberration-degraded images. $S$ is the biological image acquired by microscope, which in this study refers to images from DNN and LIVECell datasets. * indicates a convolution operation. $PSF_{\text{zernike}}$ represents the point spread function with aberration, where $PSF_{\text{zernike}}$ is generated by Zernike polynomial equation simulation. The equation can be expressed as :

$$F^{-1}(pe^{2\pi i \varphi_n/\lambda - if(m)})$$

where $F^{-1}$ denotes the 2D Fourier transform, $p$ is the amplitude of the pupil function, which is determined by detection numerical aperture (NA) and pupil coordinates. Since attenuation of amplitude is not considered in this study, we set $p = 1$. $\lambda$ as the wavelength. $f(m)$ refers to the part of the pupil equation determined by the refractive index of the immersion medium. $\varphi_n$ refers to the wavefront aberration, which is constructed by

Zernike polynomials 4-18 (Wyant ordering), and it can be expressed as:

$$\varphi_n = \sum_n a_n Z_n$$

where $Z_n$ and $a_n$ represent Zernike polynomials and amplitudes of order $n$.

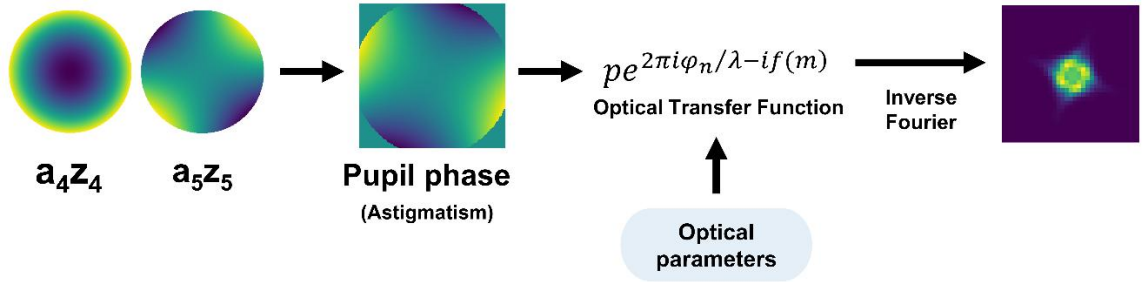

Figure 13: PSF image generation process, using the DNN dataset as an example

Figure 13 shows the PSF image generation process. We assign different amplitudes to different terms of the Zernike polynomials and combine them to simulate the wavefront aberration. By combining the above wavefront aberration with the amplitude function of the optical pupil, we generate the pupil phase, which represents the modulation of the phase of the incident light wave by the lens or optical system. Using the optical parameters and the above pupil phase, we generate the spatial frequency distribution of the optical system, i.e., the Optical Transfer Function (OTF). The OTF is subjected to an inverse Fourier transform to finally obtain a PSF image. The optical parameters used in the generation process and their physical meanings are described in Table S3 in the Supporting Information.

For single-class aberrated image generation, we selected four common aberration types, Astigmatism (set Z4 and Z5 amplitudes), Coma (set Z6 and Z7 amplitudes), Spherical (set Z8 amplitude), and Trefoil (set Z9 and Z10 amplitudes) [46], and these aberrations were added to cell images. For mixed aberrated image generation, we generated four different types of mixed aberrated image datasets, each constituting different orders of Zernike polynomials (4-6, 4-8, 4-13, and 4-16). For single-class aberrated images, we set the amplitude range as [0, 1] and sampled every 0.2 units.

Additionally, considering that images with minor magnitude aberrations are more common in practical cell imaging, we further sampled within the range [0, 0.2] with a sampling interval of 0.05. For multi-class aberrated images, we set Zernike polynomials with Gaussian sampling amplitudes ranging from 0 to 1.0 for each order.

**PLCM.** As shown in Figure 14, PLCM incorporates a pre-trained MobileNetV2 model, extracting 1280 input features from its final linear layer. Subsequently, the original classification layer of the MobileNetV2 model is replaced with three new linear layers, each designated for distinct classification tasks. The first classification head, termed Category classifier, processes 1280-dimensional input features to output predictions for 2 categories. The second head, the Aberration Type classifier, also handles 1280-dimensional input features but outputs predictions for 8 categories. The third head, the Amplitude classifier, operates on the same 1280-dimensional input and generates predictions for 9 categories. During the forward propagation, input images undergo feature extraction via MobileNetV2, and these features are simultaneously fed into the Category classifier, Aberration Type classifier, and Amplitude classifier. Each classification head produces a vector representing the predicted results for the respective classification task. In the training phase, the Mean Squared Error Loss (MSELoss) is employed to quantify the discrepancy between the predicted results and the ground truth labels. The Adam optimizer is utilized to update the model parameters. During training, the model undergoes forward and backward propagation on both the training and validation sets, computing losses and updating weights accordingly.

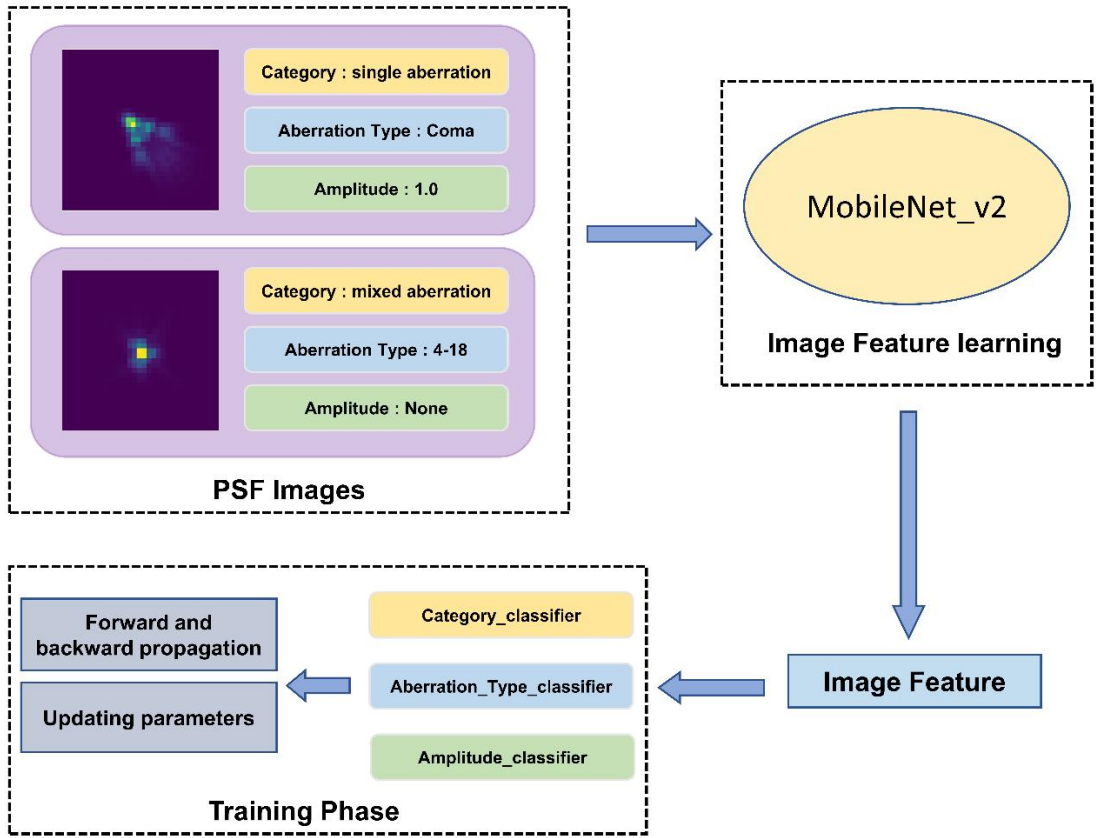

Figure 14: The architecture of PLCM.

The PSF image dataset used for PLCM training is generated by an aberration image simulator. The training dataset consists of 200 single aberration PSF images, obtained by combining four types of aberrations (Astigmatism, Coma, Spherical, and Trefoil) with eight aberration amplitudes (0.05, 0.1, 0.15, 0.2, 0.4, 0.6, 0.8, 1.0). To ensure the effectiveness of data augmentation, a specific sampling strategy is adopted in this paper. Specifically, for amplitudes greater than or equal to 0.4, fine sampling is performed at intervals of 0.01 within the range of ±0.05 of their baseline values. In this way, these sampling values within the allowable range are also considered valid amplitudes. For example, an amplitude of 0.4 can be extended to the range of 0.35 to 0.45 in practical operation. Similarly, 0.6 corresponds to the range of 0.55 to 0.65, 0.8 equates to the range of 0.75 to 0.85, and an amplitude of 1.0 is considered to cover the range of 0.95 to 1.05. This approach aims to enhance data diversity and generalization capability. Additionally, the training dataset includes 2000 mixed aberration PSF images, composed of four types of aberrations (4~6 orders, 4~8 orders, 4~13 orders, 4~18 orders) PSF images. Each type contains 500 randomly generated PSF images. For these mixed aberration PSF images, their amplitude labels are set to None. The test set includes 20 single aberration PSF images and 200 randomly generated mixed aberration PSF images.

**The Training Process**

- C3 with ResNet50 based on Mask R-CNN: Trained for 30 epochs using pre-trained weights from the official PyTorch repository, with an initial learning rate of 0.005. The batch size was set to 8.

- FPN with ResNet50 based on Mask R-CNN: Trained for 30 epochs using pre-trained weights from the official PyTorch repository, with an initial learning rate of 0.005. The batch size was set to 8.

- FPN with ResNet101 based on Mask R-CNN: Trained for 30 epochs using pre-trained weights from the official PyTorch repository, with an initial learning rate of 0.005. The learning rate was reduced by a factor of 0.2 at the 25th epoch. The batch size was set to 8.

- FPN with VGG19 based on Mask R-CNN: Trained for 15 epochs using pre-trained weights from the official PyTorch repository, with an initial learning rate of 0.005. The batch size was set to 8.

- FPN with SwinT based on Mask R-CNN: Trained for 30 epochs using pre-trained weights from the official PyTorch repository, with an initial learning rate of 0.0001. The learning rate was reduced by a factor of 0.2 at the 25th epoch. The batch size was set to 4.

- FPN with SwinS based on Mask R-CNN: Trained for 60 epochs using pre-trained weights from the official PyTorch repository, with an initial learning rate of 0.0001. The learning rate was reduced by a factor of 0.2 at the 50th epoch. The batch size was set to 4.

- MobileNet: Trained for 84 epochs using pre-trained weights from the official PyTorch repository, with an initial learning rate of 0.001. Every 20 epochs, the learning rate decays to 90% of the previous value. The batch size was set to 32.
- EfficientNet: Trained for 93 epochs using pre-trained weights from the official PyTorch repository, with an initial learning rate of 0.001. Every 20 epochs, the learning rate decays to 90% of the previous value. The batch size was set to 32.
- SqueezeNet: Trained for 120 epochs using pre-trained weights from the official PyTorch repository, with an initial learning rate of 0.001. Every 20 epochs, the learning rate decays to 90% of the previous value. The batch size was set to 32.
- DenseNet: Trained for 136 epochs using pre-trained weights from the official PyTorch repository, with an initial learning rate of 0.001. Every 20 epochs, the learning rate decays to 90% of the previous value. The batch size was set to 32.

**Supporting Information**

Comparison of cell datasets captured by fluorescence microscopy (Table S1);

Comparison of cell datasets captured by bright field microscopy (Table S2);

Optical parameter settings for DNN dataset and LIVECell dataset (Table S3);

Computational costs of the network heads and network backbones (Table S4);

The AP value comparison of different network heads (Table S5);

The AP value comparison of different network backbones (Table S6);

AP50 and AP75 value comparison of Cellpose2.0 and anchor-free model under single optical aberrations (Table S7);

AP50 and AP75 value comparison of Cellpose2.0 and anchor-free model under mixed optical aberrations (Table S8);

Morphologies of five cell types in the DNN dataset (Figure S1);

Morphologies of eight cell types in the LIVECell dataset (Figure S2);

HepG2 cells captured by bright field and fluorescence microscopy (Figure S3).

**CRediT authorship contribution statement**

Boyuan Peng: Writing – review & editing, Writing – original draft, Visualization, Validation, Methodology, Investigation. Jiaju Chen: Writing – original draft, Validation, Methodology, Investigation. P. Bilha Githinji: Visualization, Validation, Methodology, Investigation. Ijaz Gul: Writing – review & editing, Writing – original draft, Supervision. Qihui Ye: Validation, Methodology. Minjiang Chen: Methodology, Investigation. Peiwu Qin: Writing – review & editing, Investigation, Funding acquisition. Xingru Huang: Methodology, Investigation. Chenggang Yan: Funding acquisition. Dongmei Yu: Supervision, Investigation, Funding acquisition. Zhenglin Chen: Supervision, Investigation, Funding acquisition.

**Declaration of competing interest**

The authors declare that they have no known competing financial interests or personal relationships that could have appeared to influence the work reported in this paper.

**Acknowledgments**

We thank the support from the National Natural Science Foundation of China31970752,32350410397; Science, Technology, Innovation Commission of ShenzhenMunicipality, JCYJ20220530143014032, JCYJ20230807113017035, WDZC20200820173710001; Shenzhen Science and Technology Program, JCYJ20230807113017035; Shenzhen Medical Research Funds, D2301002; Department of Chemical Engineering-iBHE special cooperation joint fund project, DCE-iBHE-2022-3; Tsinghua Shenzhen International Graduate School Crossdisciplinary Research and Innovation Fund Research Plan, JC2022009; and Bureau ofPlanning, Land and Resources of Shenzhen Municipality (2022) 207.

**Data availability**

The data and code that support the findings of this study are openly available at https://github.com/Xpeng1999/Cell-Segmentation-Robustness. Other data will be made available on request.

Supporting Information

# Practical Guidelines for Cell Segmentation Models Under Optical Aberrations in Microscopy

Boyuan Peng[1,2,3#], Jiaju Chen[1,2,3#], P. Bilha Githinji[1,2,3#], Ijaz Gul[1,2,3], Qihui Ye[1,2,3], Minjiang Chen[1], Peiwu Qin[1,2,3], Xingru Huang[3], Chenggang Yan[3], Dongmei Yu[3,4]*, Jiansong Ji[1]*, Zhenglin Chen[1,2,3]*

1. Zhejiang Key Laboratory of Imaging and Interventional Medicine, Zhejiang Engineering Research Center of Interventional Medicine Engineering and Biotechnology, The Fifth Affiliated Hospital of Wenzhou Medical University, Lishui 323000, China

2. Institute of Biopharmaceutical and Health Engineering, Shenzhen International Graduate School, Tsinghua University, Shenzhen, Guangdong, China

3. School of Automation, Hangzhou Dianzi University, Hangzhou, Zhejiang Province, 310018, China

4. School of Mechanical, Electrical & Information Engineering, Shandong University, Weihai, Shandong 264209, China

#These authors contributed equally to this work.

*Corresponding authors. E-mail: yudongmei198011@sina.cn (P. Yu); jjstcty@sina.com (P. Ji); zhenglin.chen@sz.tsinghua.edu.cn (D. Chen)

## 1. Cell Dataset

To make our guideline have stronger applicability and stability, we focus on four indicators in the selection of the dataset (number of images, types of cells, image resolution and quality, cell annotations). 1) The number of images in the dataset directly affects the testing accuracy and reliability of the algorithm. The greater the number of images, the more representative the performance of the algorithm is in different situations, which can better test the robustness of the algorithm under various conditions. 2) Different types of cells may have different morphological characteristics, so the more cell types are included in the dataset, the more comprehensively the generalization ability of the algorithm can be assessed. 3) High-resolution and high-quality images capture more cellular details and help to improve the accuracy of segmentation algorithm robustness tests. 4) Whether or not the dataset is labeled with cells at the pixel level is the basis for determining whether or not the dataset can be used for the cell segmentation task, which also helps us to quickly cull out those datasets that do not belong to the cell segmentation task.

**Table S1** Comparison of cell datasets captured by fluorescence microscopy

| Dataset | Number of images | Cell types | Optical resolution (μm) | Annotation |
| --- | --- | --- | --- | --- |
| S-BSST265[1] | 79 | **10** | **0.27** | All images |
| BBBC025[2] | **138240** | 1 | 0.656 | - |
| DynamicNuclearNet[3] | 7084 | 5 | 0.37 (min) / 0.55 (max) | All images |

We selected some representative public datasets captured by fluorescence microscopy and compared their Number of images, Cell types, Optical resolution (μm), and Annotation metrics, and represented them in Table S1. The S-BSST265 dataset has the richest range of cell types and the smallest optical resolution and also has detailed pixel-level border annotation, but the number of images is only 79, which cannot meet the large amount of data required for algorithmic robustness testing. BBBC025 has 138240 images, but they do not have pixel-level border annotation, which makes it difficult to use a large amount of data images in cell segmentation. Although

DynamicNuclearNet (DNN) is not optimal in all aspects, it meets the requirements of this study. The cell types included in the DNN dataset are RAW 264.7, HEK293, HeLa-S3, NIH-3T3, and PC-3, which share similar cell morphology, so we consider these five types of cells as one class to test the robustness.

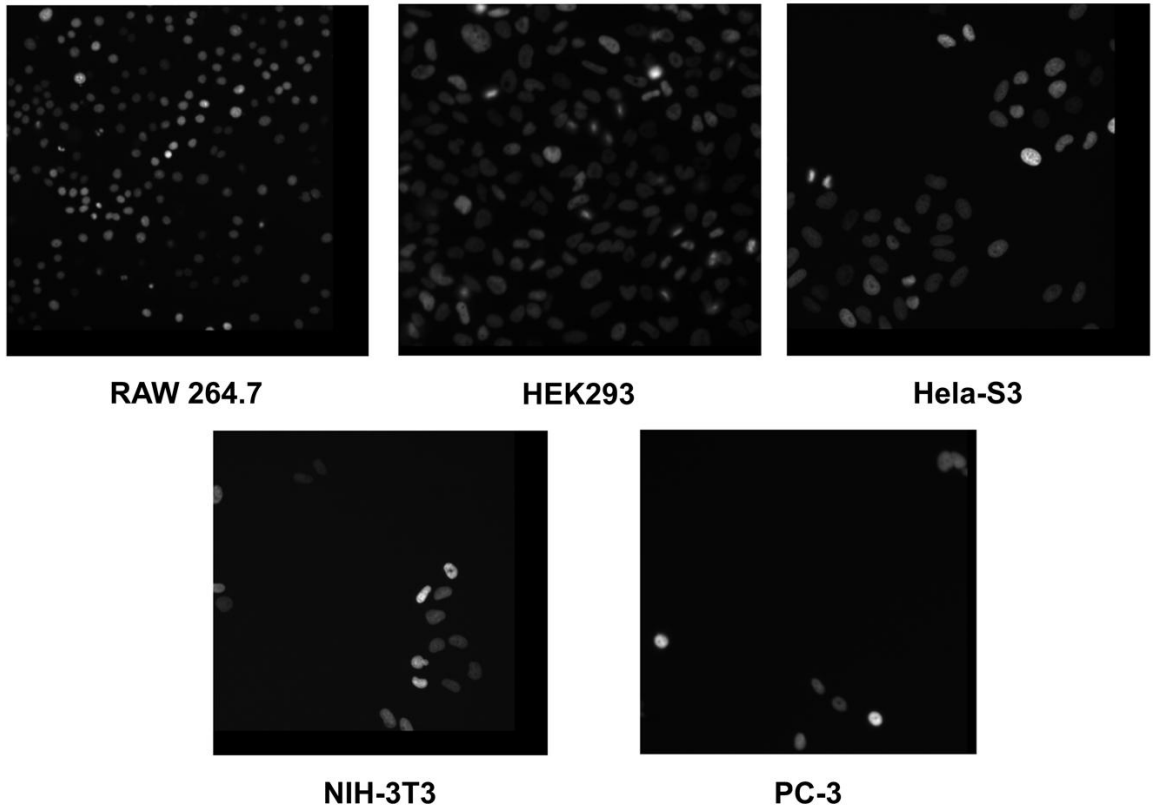

Figure S1: Morphologies of five cell types in the DNN dataset

From Figure S1, we can observe that the five types of cells in the DNN dataset have similar cell morphology, so we consider these five types of cells as one class to test the robustness.

**Table S2** Comparison of cell datasets captured by bright field microscopy

| Dataset | Number of images | Cell types | Optical resolution (μm) | Annotation |
| --- | --- | --- | --- | --- |
| BBBC041[4] | 1328 | 1 | - | All images |
| Rat astrocyte cells[5] | 1120 | 1 | **0.121 (min) / 0.242 (max)** | - |
| LIVECell[6] | **5239** | **8** | 0.61 | COCO standard |

Table S2 shows three common publicly available cell datasets captured with bright field microscopy. The LIVECell dataset is the richest in both the number of images and the types of cells, and although its optical resolution is not as good as that of Rat astrocyte cells, it is also in the range commonly used for cell imaging with bright field microscopy. Furthermore, the labeling of the LIVECell dataset adopts the COCO dataset standard [7], which makes the LIVECell dataset more user-friendly in cell segmentation tasks. Therefore, we ultimately chose the LIVECell dataset for robustness testing. Figure S2 shows the morphologies of eight cell types in the LIVECell dataset.

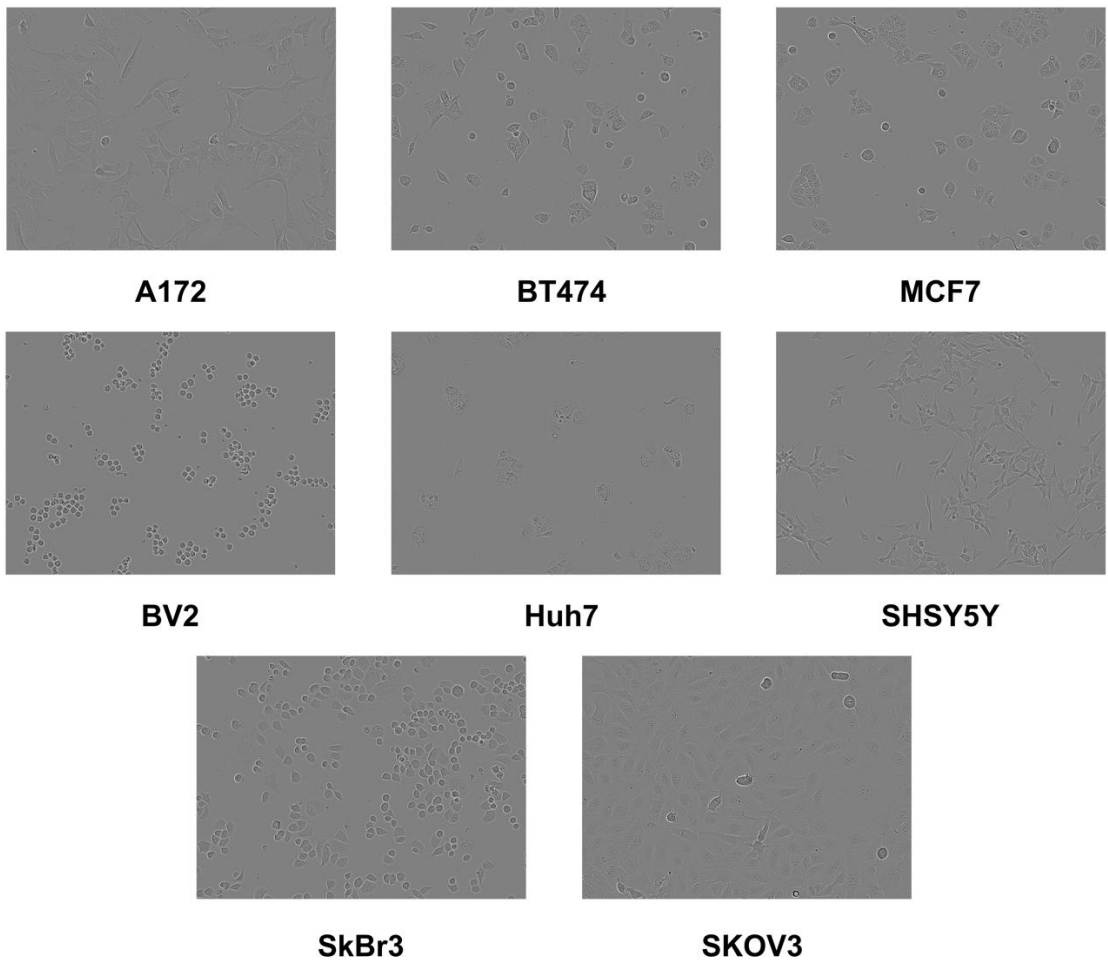

Figure S2: Morphologies of eight cell types in the LIVECell dataset

**2. Cell segmentation algorithms are chosen based on the morphological features**

It is important to note that the algorithm tested is selected based on cellular morphology rather than the device used to capture the cells. This is because cell segmentation algorithms are primarily sensitive to the morphological features of the cells rather than the imaging modality. As shown in Figure S3, even when imaging the same type of cell, the use of different capture devices, such as bright field or fluorescence microscopy, can result in variations in the appearance of the cells within the images [8]. For instance, bright-field microscopy typically provides high-contrast images that highlight the overall structure of cells, whereas fluorescence microscopy allows for the visualization of specific cellular components or proteins tagged with fluorescent markers. These differences in imaging techniques can lead to significant variations in the visual characteristics of the cells, such as edge sharpness, contrast, and the visibility of intracellular features. Consequently, the morphology of the cells as captured in the images can differ substantially depending on the modality used, which in turn affects how the segmentation algorithm interprets and processes the image data. Thus, the choice of algorithm is guided by the specific morphological features present in the images, ensuring that the segmentation process is optimized for the actual appearance of the cells rather than the characteristics of the imaging device itself.

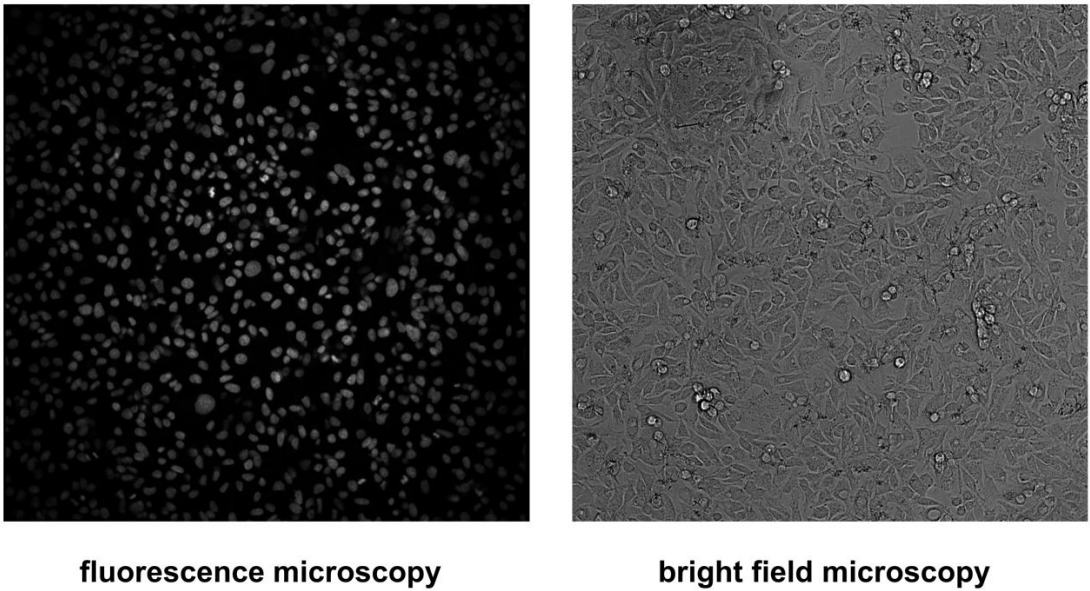

Figure S3: HepG2 cell captured by bright field and fluorescence microscopy

**3. Optical parameter settings for PSF image generation**

When generating Point Spread Function (PSF) images, the parameters of the optical system play a crucial role. These parameters define the physical characteristics of the optical system and subsequently influence the propagation of light waves and the quality of the imaging. Below are the key optical system parameters and their impact on the generation of PSF images:

Numerical Aperture (NA):

The numerical aperture is a measure of the light-gathering ability of an optical system, defined as the product of the refractive index of the lens or microscope objective and the sine of the maximum half-angle between the incident light and the optical axis. Mathematically, $NA = n * \sin(\theta)$, where n is the refractive index of the medium, and $\theta$ is the half-angle of the incident light. A larger NA allows the system to collect and focus more light, which theoretically enables the resolution of smaller features, resulting in a more concentrated PSF and higher resolution. However, an excessively large NA may also introduce stronger aberrations, negatively affecting image quality.

Wavelength (Lambda):

In microscopy, the wavelength (Lambda) represents the light's wavelength involved in image formation. In bright-field microscopy, this refers to the wavelength of the incident illumination light, which can range across the visible spectrum. In fluorescence microscopy, however, it typically refers to the wavelength of the emitted fluorescence light, which is generally longer than the excitation wavelength used to illuminate the sample. The wavelength plays a critical role in determining the system's resolution and the characteristics of the point spread function (PSF). Shorter wavelengths, whether in bright-field illumination or fluorescence emission, yield higher resolution and a more concentrated PSF. Longer wavelengths result in a more diffuse PSF. Additionally, the wavelength is essential in calculating optical path difference (OPD) and evaluating aberrations within the optical system.

Refractive Index (n):

The refractive index n indicates the ratio of the speed of light in the medium to its speed in a vacuum. The refractive index determines how light refracts and focuses within different media. The refractive index affects the propagation path of light, thereby influencing the shape and distribution of the PSF. For instance, a higher refractive index can result in tighter focusing, producing a more concentrated PSF.

Pixel Size:

Pixel size defines the physical dimensions of a single pixel in the imaging sensor, such as a CMOS or CCD. Pixel size affects the degree of discretization and sampling resolution of the image. During PSF generation, pixel size influences the level of detail in the resulting PSF image by affecting the discretization of the coordinate grid.

Medium Refractive Index (nMed):

nMed refers to the refractive index of the medium (such as the immersion liquid or medium used in microscopy). Similar to the refractive index n, nMed affects the propagation path of light, particularly at the boundaries between different media. Changes in the medium's refractive index influence the system's optical transfer function (OTF), thereby affecting the distribution of the PSF.

Zernike Polynomial Coefficients (phase Z):

Zernike polynomials are used to describe the aberrations of the optical system. Each coefficient corresponds to a specific type of aberration, such as spherical aberration or coma. These coefficients directly affect the shape of the PSF. Non-zero Zernike coefficients can introduce distortions or blur in the center of the PSF, reflecting the aberrations present in the actual optical system.

**Table S3:** Optical parameter settings for DNN dataset and LIVECell dataset

| Dataset | NA | Lambda (μm) | n | Pixel Size (μm) | nMed |
|---|---|---|---|---|---|
| DNN | 0.75 | 0.35 | 1 | 0.25 | 1 |
| LIVECell | 1.35 | 0.55 | 1 | 1.24 | 1 |

Table S3 shows the above optical parameter settings for the DNN dataset and LIVECell dataset.

## 4. PSF image acquisition protocol

Users can obtain the point spread function of the microscope system through the following experimental steps[9]:

Step 1: Prepare the Sample

Use very small fluorescent beads as point sources.

1. Select Appropriate Fluorescent Beads:

• Ensure the fluorescence spectrum of the beads matches the excitation light and detection channels of the microscope.

• Commonly used bead diameters range from 20 to 200 nanometers.

2. Prepare the Sample:

• Dispense a dilute solution of fluorescent beads onto a microscope slide, ensuring sparse distribution so only a few beads are visible in each field of view.

• Cover with a coverslip and secure with nail polish or another sealing agent.

Step 2: Adjust the Microscope

Ensure the microscope system is in optimal working condition, including proper calibration and alignment.

1. Calibrate the Microscope:

• Perform calibration to ensure all optical components are correctly positioned.

• Verify that the focus system and objectives are in optimal condition.

2. Set Parameters:

• Adjust the excitation light intensity and exposure time to achieve high signal-to-noise ratio images.

• Use a high numerical aperture (NA) objective lens to increase resolution.

Step 3: Acquire Images

Capture images of the fluorescent beads with the microscope, ensuring clear PSF images.

1. Locate Beads:

• Find single fluorescent beads under the microscope and center them in the field of view.

• Ensure beads are sparsely distributed to avoid overlap.

2. Capture Images:

• Adjust the focus to achieve the clearest image of the beads, minimizing blur.

• Capture multiple images of the beads to ensure accuracy and reproducibility.

The above steps will be included in the supplementary information.

**5. Computational costs of the network heads and network backbones**

**Table S4:** Computational costs of the network heads and network backbones

| Backbone+Head | Computational complexity (FLOPs) | Total number of model parameters (Params) |
|---|---|---|
| Resnet50+C3 | 118.410 GFLOPs | 74.83 M |
| Resnet50+FPN | 134.486 GFLOPs | 43.97 M |
| Resnet101+FPN | 132.136 GFLOPs | 62.31 M |
| Vgg19+FPN | 280.842 GFLOPs | 39.32 M |
| SwinT+FPN | 110.783 GFLOPs | 29.25 M |
| SwinS+FPN | 146.265 GFLOPs | 43.44 M |

The results of Table S4 show that the SwinS model achieved the best segmentation performance for this task, albeit with a relatively high computational cost (FLOPs: 146.265 GFLOPs, Params: 43.44 M). On the other hand, the SwinT model, while slightly less accurate than SwinS, significantly reduced computational overhead (FLOPs: 110.783 GFLOPs, Params: 29.25 M) and performed well in resource-constrained or real-time scenarios. To optimize these algorithms for practical use, we suggest employing model compression techniques such as quantization, pruning, or distillation to reduce model size and computation while maintaining accuracy. Additionally, users can choose between SwinS and SwinT based on specific needs: SwinS is ideal when computational resources are abundant and high segmentation accuracy is crucial, while SwinT offers a more economical and efficient alternative in scenarios with limited resources or stringent real-time requirements. This optimization strategy enhances the efficiency of segmentation tasks and provides more flexible solutions for practical applications.

**6. The AP value of different cell instance segmentation models**

Table S5: The AP value comparison of different network heads.

| | | C3 | FPN |
|---|---|---|---|
| Original image | | 39.2729203 | 41.44031859 |
| Astigmatism | 0.05 | 34.95595951 | 37.95526735 |
| | 0.1 | 34.19368284 | 37.39523426 |
| | 0.15 | 33.50788561 | 36.16724681 |
| | 0.2 | 33.10851888 | 34.38587681 |
| | 0.4 | 30.47711126 | 30.80669065 |
| | 0.6 | 27.37686249 | 30.69583761 |
| | 0.8 | 26.014191 | 29.78061269 |
| | 1.0 | 24.5637096 | 28.70913177 |
| Coma | 0.05 | 34.65787951 | 38.15049872 |
| | 0.1 | 34.57738429 | 37.52688349 |
| | 0.15 | 33.66855431 | 36.83931276 |
| | 0.2 | 33.00338939 | 35.60677248 |
| | 0.4 | 30.24066687 | 30.25868021 |
| | 0.6 | 28.25900576 | 28.75017273 |
| | 0.8 | 26.49208968 | 27.62840318 |
| | 1.0 | 24.17305661 | 26.62680382 |
| Spherical | 0.05 | 34.99734822 | 38.17104621 |
| | 0.1 | 34.30367937 | 37.70954927 |
| | 0.15 | 33.99980416 | 36.71538078 |
| | 0.2 | 33.31639491 | 35.56041235 |
| | 0.4 | 29.855605 | 30.65107584 |
| | 0.6 | 25.28309659 | 29.82684033 |
| | 0.8 | 23.0296269 | 28.48111981 |
| | 1.0 | 18.42619756 | 27.0719311 |
| Trefoil | 0.05 | 34.99734822 | 38.00804211 |
| | 0.1 | 34.37351301 | 37.60653904 |
| | 0.15 | 34.12469706 | 36.4029914 |
| | 0.2 | 33.25774569 | 35.0581903 |
| | 0.4 | 31.30157847 | 31.10745741 |
| | 0.6 | 27.91996455 | 30.58599596 |
| | 0.8 | 25.96926944 | 29.79819906 |
| | 1.0 | 24.57655793 | 29.06553865 |
| Mixed | 4-6 | 33.09573629 | 35.70374052 |
| | 4-8 | 31.53377591 | 32.26522159 |
| | 4-13 | 31.77766193 | 32.44395259 |
| | 4-18 | 32.2514844 | 32.8570048 |

Table S6: The AP value comparison of different network backbones, the baseline is the Otsu thresholding method.

| | | ResNet50 | ResNet101 | VGG19 | SwinT | SwinS | Otsu |
|---|---|---|---|---|---|---|---|
| Original image | | 41.44031859 | 40.588069 | 40.9120008 | 42.83962057 | 44.2704675 | 17.7885222 |
| Astigmatism | 0.05 | 37.95526735 | 39.43746495 | 39.7927479 | 42.15685756 | 43.6427549 | 15.1663947 |
| | 0.1 | 37.39523426 | 39.60611466 | 39.8780371 | 41.89651926 | 43.5977675 | 15.1460810 |
| | 0.15 | 36.16724681 | 39.32513467 | 39.7852854 | 41.78674795 | 43.4696165 | 14.8577201 |
| | 0.2 | 34.38587681 | 38.66964053 | 39.5471267 | 41.71641141 | 43.2392693 | 14.6231688 |
| | 0.4 | 30.80669065 | 38.06798666 | 39.2346021 | 38.48436835 | 41.6646844 | 13.1064104 |
| | 0.6 | 30.69583761 | 38.07474454 | 39.0796943 | 36.62089164 | 39.7589657 | 11.8945282 |
| | 0.8 | 29.78061269 | 37.45294134 | 38.7059676 | 36.44445203 | 38.2666231 | 11.4583340 |
| | 1.0 | 28.70913177 | 37.14913666 | 38.5524319 | 36.02941547 | 36.55708 | 11.4180617 |
| Coma | 0.05 | 38.15049872 | 39.73290822 | 39.9365838 | 41.94467718 | 43.6490844 | 15.0672780 |
| | 0.1 | 37.52688349 | 39.12091699 | 39.7210862 | 42.07634932 | 43.5721692 | 14.7041544 |
| | 0.15 | 36.83931276 | 39.22779115 | 39.7971119 | 41.75474895 | 43.4769993 | 14.3820530 |
| | 0.2 | 35.60677248 | 39.00141702 | 39.7003696 | 41.49566265 | 43.2635261 | 14.0036472 |
| | 0.4 | 30.25868021 | 38.09778683 | 39.24734 | 38.65116174 | 41.2739913 | 11.3338773 |
| | 0.6 | 28.75017273 | 37.61674586 | 39.0566047 | 37.52975397 | 38.6489474 | 10.9040295 |
| | 0.8 | 27.62840318 | 37.7251046 | 39.1218809 | 37.31052136 | 37.1928445 | 10.4013815 |
| | 1.0 | 26.62680382 | 37.31830717 | 38.8359525 | 37.51327693 | 36.1964315 | 10.3356756 |
| Spherical | 0.05 | 38.17104621 | 39.76957038 | 39.94342 | 42.17007899 | 43.6610814 | 15.9764575 |
| | 0.1 | 37.70954927 | 39.51112795 | 39.8774193 | 41.91804453 | 43.6012477 | 15.4577844 |
| | 0.15 | 36.71538078 | 38.81119154 | 39.6421758 | 42.03085586 | 43.4730882 | 15.0181917 |
| | 0.2 | 35.56041235 | 38.55833189 | 39.548345 | 41.61499365 | 43.2445661 | 14.5552505 |
| | 0.4 | 30.65107584 | 38.11980707 | 39.1871241 | 39.12850478 | 40.6639125 | 11.5156772 |
| | 0.6 | 29.82684033 | 37.75877642 | 38.8236842 | 37.21069439 | 38.2125434 | 11.1483271 |
| | 0.8 | 28.48111981 | 37.57360767 | 38.5633739 | 37.01668534 | 37.6585296 | 11.1188444 |
| | 1.0 | 27.0719311 | 37.77620714 | 38.3660956 | 37.03436777 | 36.7743986 | 10.5834335 |
| Trefoil | 0.05 | 38.00804211 | 39.43409968 | 39.8013934 | 42.17007899 | 43.6510519 | 16.1456677 |
| | 0.1 | 37.60653904 | 39.59128611 | 39.9104315 | 41.91804453 | 43.6103852 | 14.5371371 |
| | 0.15 | 36.4029914 | 38.96494801 | 39.7216298 | 42.03085586 | 43.493776 | 13.5008468 |
| | 0.2 | 35.0581903 | 39.10914881 | 39.8313642 | 41.61499365 | 43.3326191 | 13.1922021 |
| | 0.4 | 31.10745741 | 38.23390308 | 39.3567683 | 39.12850478 | 41.6995808 | 12.1462478 |
| | 0.6 | 30.58599596 | 38.17387245 | 39.1237874 | 37.21069439 | 39.5650642 | 12.1062238 |
| | 0.8 | 29.79819906 | 37.87644907 | 38.9519508 | 37.01668534 | 38.2343119 | 11.3987050 |
| | 1.0 | 29.06553865 | 37.70190732 | 38.8667551 | 37.03436777 | 37.4000048 | 11.1505284 |
| Mixed | 4-6 | 35.70374052 | 38.98892639 | 39.6701387 | 41.5665665 | 43.3138401 | 14.8501710 |
| | 4-8 | 32.26522159 | 38.72627261 | 39.5608628 | 40.03848815 | 42.482869 | 13.3286309 |
| | 4-13 | 32.44395259 | 38.3285 199 | 39.4122647 | 40.1273194 | 42.4321965 | 13.0331464 |
| | 4-18 | 32.8570048 | 38.47661737 | 39.4309549 | 40.84878921 | 42.6935946 | 12.1671275 |

**Table S7:** AP50 and AP75 value comparison of Cellpose2.0 and anchor-free model under single optical aberrations.

| | AP50 | AP75 |
|---|---|---|

|   |   |   | Cellpose2.0 | anchor-free | Cellpose2.0 | anchor-free |
|---|---|---|---|---|---|---|
| A172 | raw image | 0 | 69.69126 | 72.1116 | 42.01347 | 30.1491 |
|   | Astigmatism | 0.05 | 69.47376 | 73.8109 | 41.103595 | 29.2828 |
|   |   | 0.1 | 69.399434 | 73.8595 | 40.950444 | 29.1114 |
|   |   | 0.15 | 69.39377 | 73.6813 | 40.79391 | 28.6221 |
|   |   | 0.2 | 69.27935 | 73.0321 | 40.615472 | 28.0692 |
|   |   | 0.4 | 65.417534 | 66.206 | 34.12319 | 19.4986 |
|   |   | 0.6 | 52.28484 | 51.2845 | 19.369052 | 8.7393 |
|   |   | 0.8 | 36.854142 | 41.5173 | 9.114588 | 3.6806 |
|   |   | 1.0 | 26.865608 | 30.5739 | 4.0050755 | 1.1989 |
|   | Coma | 0.05 | 69.50606 | 73.8029 | 41.036606 | 29.2152 |
|   |   | 0.1 | 69.36503 | 73.8454 | 41.04656 | 29.1155 |
|   |   | 0.15 | 69.35153 | 73.7346 | 40.963425 | 28.6697 |
|   |   | 0.2 | 69.38353 | 73.2803 | 40.74859 | 28.0292 |
|   |   | 0.4 | 67.14087 | 67.2323 | 37.07587 | 21.6984 |
|   |   | 0.6 | 60.23118 | 55.0162 | 24.60264 | 12.2677 |
|   |   | 0.8 | 50.958538 | 45.4415 | 15.682495 | 6.5716 |
|   |   | 1.0 | 39.854088 | 35.598 | 7.893492 | 2.7701 |
|   | Spherical | 0.05 | 69.49943 | 73.8213 | 41.105534 | 29.2673 |
|   |   | 0.1 | 69.5056 | 73.9558 | 40.98519 | 29.1283 |
|   |   | 0.15 | 69.56369 | 73.8772 | 41.016953 | 28.9534 |
|   |   | 0.2 | 69.53162 | 73.4075 | 40.849895 | 28.3619 |
|   |   | 0.4 | 66.498634 | 65.4185 | 35.25383 | 19.1715 |
|   |   | 0.6 | 46.515144 | 36.2734 | 12.516299 | 4.5774 |
|   |   | 0.8 | 48.214832 | 38.3603 | 12.082213 | 4.4761 |
|   |   | 1.0 | 44.701225 | 34.4805 | 9.943868 | 4.2974 |
|   | Trefoil | 0.05 | 69.42837 | 73.8007 | 41.17365 | 29.3057 |
|   |   | 0.1 | 69.44662 | 73.8474 | 41.081272 | 29.1338 |
|   |   | 0.15 | 69.42781 | 73.8316 | 41.05466 | 29.0029 |
|   |   | 0.2 | 69.33746 | 73.4935 | 40.740776 | 28.4052 |
|   |   | 0.4 | 67.70922 | 68.9615 | 37.547794 | 22.5913 |
|   |   | 0.6 | 61.0161 | 60.9793 | 26.454077 | 14.7863 |
|   |   | 0.8 | 55.043274 | 53.6034 | 18.798048 | 10.2462 |
|   |   | 1.0 | 54.464264 | 48.8605 | 16.832294 | 8.3697 |
| BT474 | raw image | 0 | 62.37256 | 77.5985 | 41.003506 | 42.5394 |
|   | Astigmatism | 0.05 | 62.123425 | 75.0519 | 38.683014 | 38.7978 |
|   |   | 0.1 | 62.01793 | 74.8071 | 38.382156 | 38.4341 |
|   |   | 0.15 | 61.632626 | 74.304 | 38.038246 | 37.8598 |
| | | 0.2 | 60.81609 | 73.1933 | 37.428005 | 36.8974 |
| | | 0.4 | 45.280437 | 62.4307 | 27.1189 | 27.2808 |
| | | 0.6 | 25.057194 | 45.5028 | 10.387698 | 16.817 |
| | | 0.8 | 19.745687 | 35.7286 | 4.061364 | 9.3872 |
| | | 1.0 | 15.547708 | 26.9966 | 1.7079221 | 3.0837 |
| | Coma | 0.05 | 62.136406 | 75.1184 | 38.732155 | 38.7452 |
| | | 0.1 | 62.13625 | 74.9156 | 38.561172 | 38.66 |
| | | 0.15 | 61.54474 | 74.5483 | 38.20104 | 38.3233 |
| | | 0.2 | 60.901047 | 73.8782 | 37.89335 | 37.574 |
| | | 0.4 | 49.335358 | 64.3823 | 30.942722 | 29.6115 |
| | | 0.6 | 30.67992 | 49.5251 | 13.036373 | 19.0052 |
| | | 0.8 | 23.67786 | 40.5404 | 7.241633 | 10.5889 |
| | | 1.0 | 17.586214 | 32.7722 | 3.7525098 | 4.0159 |
| | Spherical | 0.05 | 62.118267 | 75.0611 | 38.73188 | 38.7566 |
| | | 0.1 | 62.08154 | 74.8908 | 38.51691 | 38.5975 |
| | | 0.15 | 61.72822 | 74.5669 | 38.320583 | 38.2475 |
| | | 0.2 | 61.591118 | 73.9913 | 38.031666 | 37.4821 |
| | | 0.4 | 48.976025 | 61.4293 | 28.919834 | 27.005 |
| | | 0.6 | 22.829828 | 36.4519 | 5.7456284 | 13.4202 |
| | | 0.8 | 21.449411 | 29.1606 | 4.75372 | 8.4348 |
| | | 1.0 | 19.75445 | 25.3105 | 4.6179595 | 7.1644 |
| | Trefoil | 0.05 | 62.249012 | 75.1426 | 38.79216 | 38.729 |
| | | 0.1 | 62.094868 | 74.9096 | 38.472675 | 38.5614 |
| | | 0.15 | 61.87103 | 74.4386 | 38.434814 | 38.0768 |
| | | 0.2 | 61.338024 | 73.7429 | 37.94525 | 37.2487 |
| | | 0.4 | 50.4161 | 64.7006 | 31.367296 | 29.13 |
| | | 0.6 | 33.79637 | 52.0969 | 15.042327 | 21.4571 |
| | | 0.8 | 25.684494 | 44.6082 | 8.094766 | 17.3461 |
| | | 1.0 | 23.90945 | 39.9273 | 7.605587 | 14.3654 |
| BV2 | raw image | 0 | 74.02836 | 84.8842 | 58.095013 | 45.2105 |
| | Astigmatism | 0.05 | 64.24896 | 83.5023 | 47.738113 | 32.5459 |
| | | 0.1 | 63.403763 | 83.3818 | 47.08164 | 31.6149 |
| | | 0.15 | 61.70879 | 82.5432 | 46.004276 | 30.0819 |
| | | 0.2 | 59.309216 | 82.2512 | 43.944954 | 27.7409 |
| | | 0.4 | 40.97708 | 77.6844 | 26.82407 | 15.7912 |
| | | 0.6 | 20.067184 | 69.8168 | 6.994044 | 5.1794 |
| | | 0.8 | 10.695336 | 56.6972 | 1.1561881 | 1.2524 |
| | | 1.0 | 4.730028 | 32.5533 | 0.09416657 | 0.373 |
| | Coma | 0.05 | 64.33093 | 83.487 | 47.794247 | 32.4715 |
| | | 0.1 | 63.862755 | 83.3921 | 47.375042 | 31.5508 |
| | | 0.15 | 62.52246 | 82.5722 | 46.632137 | 30.1056 |
| | | 0.2 | 60.85167 | 82.3496 | 45.511185 | 28.2381 |
| | | 0.4 | 47.3087 | 78.3602 | 35.50274 | 18.6009 |
| | | 0.6 | 32.471943 | 72.6692 | 13.586995 | 7.0237 |
| | | 0.8 | 19.734793 | 62.8036 | 4.064562 | 0.9201 |
| | | 1.0 | 10.724328 | 41.3667 | 0.6729487 | 0.0305 |
| | Spherical | 0.05 | 64.34633 | 83.5053 | 47.752213 | 32.5404 |
| | | 0.1 | 63.925205 | 83.4231 | 47.478024 | 31.5735 |
| | | 0.15 | 62.78274 | 82.5855 | 46.759575 | 29.9817 |
| | | 0.2 | 61.429913 | 82.3575 | 45.587833 | 27.4323 |
| | | 0.4 | 42.127594 | 76.7316 | 27.324963 | 9.8364 |
| | | 0.6 | 15.382467 | 62.0906 | 1.9030633 | 0.904 |
| | | 0.8 | 18.635563 | 52.8182 | 1.7137939 | 1.2169 |
| | | 1.0 | 17.607697 | 46.4468 | 2.4859483 | 0.3757 |
| | Trefoil | 0.05 | 64.40643 | 83.5088 | 47.966347 | 32.6357 |
| | | 0.1 | 63.932545 | 83.4312 | 47.57596 | 31.7041 |
| | | 0.15 | 62.89484 | 83.1709 | 46.872272 | 30.2543 |
| | | 0.2 | 61.389805 | 82.3838 | 45.663166 | 28.379 |
| | | 0.4 | 48.53557 | 78.9674 | 34.14077 | 16.2345 |
| | | 0.6 | 26.255241 | 73.4379 | 11.980423 | 6.0418 |
| | | 0.8 | 17.1408 | 68.3029 | 4.0656133 | 1.8183 |
| | | 1.0 | 16.128796 | 63.3395 | 3.8666363 | 0.8994 |
| Huh7 | raw image | 0 | 71.84867 | 78.7316 | 53.157322 | 50.375 |
| | Astigmatism | 0.05 | 71.27176 | 76.9253 | 50.8927 | 45.6376 |
| | | 0.1 | 71.12192 | 76.5886 | 50.797375 | 44.9276 |
| | | 0.15 | 70.60881 | 75.9882 | 49.958614 | 44.2019 |
| | | 0.2 | 69.450035 | 74.8226 | 48.400757 | 41.7685 |
| | | 0.4 | 45.736496 | 54.2555 | 23.25953 | 18.3593 |
| | | 0.6 | 23.983847 | 32.3299 | 6.0870705 | 5.6982 |
| | | 0.8 | 19.479824 | 24.8959 | 3.3703392 | 2.7195 |
| | | 1.0 | 18.603212 | 21.9847 | 2.3710675 | 1.0673 |
| | Coma | 0.05 | 71.29161 | 77.002 | 50.925392 | 45.632 |
| | | 0.1 | 71.15416 | 76.8034 | 50.90866 | 45.1783 |
| | | 0.15 | 70.749435 | 76.236 | 50.232227 | 44.6355 |
| | | 0.2 | 69.793884 | 75.441 | 49.1481 | 42.937 |
| | | 0.4 | 50.419586 | 56.4854 | 28.084898 | 20.6481 |
| | | 0.6 | 33.129906 | 36.5117 | 11.744976 | 8.2727 |
| | | 0.8 | 24.408886 | 29.3182 | 6.720874 | 4.2951 |
|  |  | 1.0 | 16.67561 | 23.6835 | 3.3201423 | 2.5509 |
|  | Spherical | 0.05 | 71.32897 | 76.8958 | 50.780754 | 45.7573 |
|  |  | 0.1 | 71.1146 | 76.8243 | 50.832237 | 45.1439 |
|  |  | 0.15 | 70.85353 | 76.2943 | 50.331055 | 44.8613 |
|  |  | 0.2 | 70.159195 | 75.2035 | 49.499275 | 42.5965 |
|  |  | 0.4 | 49.411747 | 50.2176 | 25.115227 | 15.0982 |
|  |  | 0.6 | 23.663996 | 26.3817 | 4.922513 | 2.5958 |
|  |  | 0.8 | 22.881893 | 27.4918 | 5.1004677 | 2.6562 |
|  |  | 1.0 | 21.09651 | 25.4389 | 4.7897654 | 3.0985 |
|  | Trefoil | 0.05 | 71.306366 | 76.9161 | 51.129604 | 45.8281 |
|  |  | 0.1 | 70.98944 | 76.7454 | 50.670372 | 45.0352 |
|  |  | 0.15 | 70.63787 | 76.1549 | 50.383682 | 44.2466 |
|  |  | 0.2 | 70.35971 | 75.0332 | 49.598637 | 42.0403 |
|  |  | 0.4 | 55.50267 | 60.7684 | 32.354748 | 22.7774 |
|  |  | 0.6 | 34.478783 | 38.6906 | 12.637788 | 8.8447 |
|  |  | 0.8 | 28.171904 | 31.7995 | 7.6523237 | 5.3619 |
|  |  | 1.0 | 25.521528 | 28.9714 | 7.01448 | 4.5379 |
| MCF7 | raw image | 0 | 57.7619 | 76.1826 | 35.91879 | 32.9693 |
|  | Astigmatism | 0.05 | 56.12194 | 72.7392 | 31.254572 | 26.8182 |
|  |  | 0.1 | 55.864304 | 71.9718 | 30.966278 | 26.5983 |
|  |  | 0.15 | 55.44922 | 71.3633 | 30.41677 | 25.5749 |
|  |  | 0.2 | 54.682518 | 69.9406 | 29.674446 | 24.3359 |
|  |  | 0.4 | 40.212463 | 54.7228 | 19.510641 | 13.8727 |
|  |  | 0.6 | 18.229713 | 30.5549 | 6.6793995 | 3.8451 |
|  |  | 0.8 | 10.485144 | 17.3384 | 2.3975613 | 1.0807 |
|  |  | 1.0 | 6.6129766 | 8.0448 | 0.85675067 | 0.1821 |
|  | Coma | 0.05 | 45.00071 | 72.7502 | 14.334236 | 26.9584 |
|  |  | 0.1 | 44.646137 | 72.4241 | 13.960932 | 26.737 |
|  |  | 0.15 | 44.006813 | 71.5449 | 13.5523815 | 26.0830 |
|  |  | 0.2 | 42.979164 | 70.2831 | 12.831484 | 25.1964 |
|  |  | 0.4 | 30.407337 | 58.2175 | 6.395548 | 17.4525 |
|  |  | 0.6 | 13.890071 | 37.3934 | 1.7438921 | 7.1036 |
|  |  | 0.8 | 7.0071774 | 25.005 | 0.39057034 | 2.5781 |
|  |  | 1.0 | 4.316065 | 15.9578 | 0.09733833 | 0.6063 |
|  | Spherical | 0.05 | 45.054314 | 72.7639 | 14.31999 | 26.9328 |
|  |  | 0.1 | 44.98696 | 72.5256 | 14.201583 | 26.7657 |
|  |  | 0.15 | 44.794838 | 71.7177 | 14.022252 | 26.1576 |
|  |  | 0.2 | 44.432343 | 70.6382 | 13.592614 | 25.2392 |
|  |  | 0.4 | 34.84883 | 56.8236 | 7.6500244 | 14.6651 |
|  |  | 0.6 | 11.067788 | 17.3195 | 0.91046876 | 2.2711 |
|  |  | 0.8 | 10.69451 | 13.4563 | 0.8506638 | 1.8779 |
|  |  | 1.0 | 10.055914 | 9.8172 | 1.0146351 | 1.9292 |
|  | Trefoil | 0.05 | 45.060875 | 72.8255 | 14.3658 | 27.0062 |
|  |  | 0.1 | 44.894955 | 72.5928 | 14.229303 | 26.8906 |
|  |  | 0.15 | 44.516296 | 71.7468 | 13.958146 | 26.1808 |
|  |  | 0.2 | 43.94043 | 70.5866 | 13.487194 | 25.2721 |
|  |  | 0.4 | 36.660576 | 60.8184 | 8.857303 | 16.8518 |
|  |  | 0.6 | 23.58351 | 29.7831 | 4.0978146 | 4.5601 |
|  |  | 0.8 | 16.700645 | 29.7831 | 2.4396265 | 4.5601 |
|  |  | 1.0 | 16.576807 | 22.8287 | 2.3931422 | 3.3893 |
| SkBr3 | raw image | 0 | 91.295296 | 90.25 | 77.71361 | 80.5415 |
|  | Astigmatism | 0.05 | 90.501976 | 90.0207 | 71.81754 | 75.6261 |
|  |  | 0.1 | 90.39946 | 89.9916 | 71.56109 | 75.5419 |
|  |  | 0.15 | 90.226074 | 89.943 | 70.976654 | 75.4291 |
|  |  | 0.2 | 89.9836 | 89.863 | 69.70617 | 74.5784 |
|  |  | 0.4 | 85.93431 | 87.7874 | 54.236263 | 68.7238 |
|  |  | 0.6 | 75.91278 | 83.7007 | 27.134878 | 51.9638 |
|  |  | 0.8 | 63.1922 | 75.5807 | 10.159981 | 28.1681 |
|  |  | 1.0 | 44.335773 | 61.1888 | 4.0683746 | 9.5794 |
|  | Coma | 0.05 | 90.51113 | 90.0108 | 71.8251 | 75.6191 |
|  |  | 0.1 | 90.4378 | 89.9903 | 71.72358 | 75.5912 |
|  |  | 0.15 | 90.31946 | 89.9664 | 71.63174 | 75.6569 |
|  |  | 0.2 | 90.17475 | 89.9052 | 71.23778 | 75.6266 |
|  |  | 0.4 | 87.38454 | 87.751 | 63.18205 | 73.5122 |
|  |  | 0.6 | 79.24119 | 83.8367 | 29.460331 | 65.3208 |
|  |  | 0.8 | 68.17987 | 77.7049 | 15.051283 | 46.8105 |
|  |  | 1.0 | 52.91324 | 69.9644 | 6.8507714 | 19.7434 |
|  | Spherical | 0.05 | 90.55651 | 90.0114 | 71.79829 | 75.6356 |
|  |  | 0.1 | 90.46536 | 89.9929 | 71.765724 | 75.646 |
|  |  | 0.15 | 90.345894 | 89.9699 | 71.619316 | 75.5762 |
|  |  | 0.2 | 90.25567 | 89.8839 | 71.30091 | 75.5057 |
|  |  | 0.4 | 86.3652 | 87.5336 | 53.921722 | 69.6875 |
|  |  | 0.6 | 68.31258 | 78.852 | 11.920143 | 44.8876 |
|  |  | 0.8 | 62.98209 | 72.7214 | 9.267128 | 37.9149 |
|  |  | 1.0 | 55.63268 | 71.3567 | 9.6744995 | 36.7509 |
|  | Trefoil | 0.05 | 90.5261 | 90.0152 | 71.91888 | 75.635 |
|  |  | 0.1 | 90.476295 | 89.9822 | 71.7862 | 75.628 |
|  |  | 0.15 | 90.38908 | 89.9615 | 71.49199 | 75.5904 |
| | | 0.2 | 90.227135 | 89.8965 | 70.894165 | 75.3017 |
| | | 0.4 | 87.80849 | 87.9125 | 61.144882 | 70.8627 |
| | | 0.6 | 81.017075 | 85.4909 | 34.522587 | 64.3958 |
| | | 0.8 | 72.4279 | 82.6978 | 16.552105 | 57.1822 |
| | | 1.0 | 65.505226 | 80.1596 | 13.998949 | 50.5459 |
| SKOV3 | raw image | 0 | 83.64671 | 86.0056 | 59.81653 | 53.0687 |
| | Astigmatism | 0.05 | 83.75906 | 84.6308 | 58.9299 | 48.7066 |
| | | 0.1 | 83.80488 | 84.3873 | 58.858627 | 47.769 |
| | | 0.15 | 83.85787 | 83.4813 | 58.652298 | 46.0697 |
| | | 0.2 | 83.61412 | 81.9867 | 58.25914 | 42.5977 |
| | | 0.4 | 79.01078 | 55.3223 | 46.564636 | 14.5455 |
| | | 0.6 | 52.93346 | 25.6469 | 14.092259 | 3.4483 |
| | | 0.8 | 36.427406 | 16.7761 | 6.3196464 | 1.5709 |
| | | 1.0 | 27.745815 | 10.7806 | 3.1491268 | 0.467 |
| | Coma | 0.05 | 83.76723 | 84.67 | 59.00312 | 48.823 |
| | | 0.1 | 83.73437 | 84.4033 | 58.894318 | 48.0862 |
| | | 0.15 | 83.7437 | 83.5468 | 58.651524 | 46.4811 |
| | | 0.2 | 83.68406 | 82.1508 | 58.52397 | 43.536 |
| | | 0.4 | 80.158455 | 59.6683 | 49.821846 | 18.2909 |
| | | 0.6 | 64.51229 | 30.7387 | 21.71158 | 5.2874 |
| | | 0.8 | 52.20173 | 20.9118 | 11.823788 | 2.476 |
| | | 1.0 | 40.913548 | 13.7732 | 6.219238 | 1.0112 |
| | Spherical | 0.05 | 83.837555 | 84.6597 | 58.925915 | 48.8281 |
| | | 0.1 | 83.842224 | 84.4443 | 58.907352 | 48.1372 |
| | | 0.15 | 83.90839 | 83.8186 | 58.784878 | 46.3515 |
| | | 0.2 | 83.71862 | 82.1914 | 58.3788 | 43.3354 |
| | | 0.4 | 78.175095 | 49.9827 | 43.63491 | 12.0379 |
| | | 0.6 | 37.97289 | 14.2329 | 6.382744 | 1.9811 |
| | | 0.8 | 41.874084 | 13.9796 | 6.392648 | 1.6494 |
| | | 1.0 | 37.302288 | 11.503 | 5.021895 | 1.5877 |
| | Trefoil | 0.05 | 83.87217 | 84.6666 | 59.056675 | 48.8429 |
| | | 0.1 | 83.81987 | 84.4017 | 58.99333 | 48.1629 |
| | | 0.15 | 83.72655 | 83.9638 | 59.13583 | 46.7451 |
| | | 0.2 | 83.611176 | 82.8502 | 58.672302 | 44.1489 |
| | | 0.4 | 81.28025 | 64.2158 | 51.918716 | 20.2337 |
| | | 0.6 | 63.37766 | 32.9207 | 21.804722 | 5.4367 |
| | | 0.8 | 50.041866 | 22.6599 | 11.450906 | 2.7336 |
| | | 1.0 | 46.07571 | 18.6643 | 8.997091 | 2.0241 |
| | raw image | 0 | 47.90515 | 61.0518 | 18.727745 | 13.8749 |
| SHSY5Y | Astigmatism | 0.05 | 45.00071 | 60.953 | 14.334236 | 10.5178 |
| | | 0.1 | 44.646137 | 60.8443 | 13.960932 | 10.3074 |
| | | 0.15 | 44.006813 | 60.2711 | 13.5523815 | 10.0408 |
| | | 0.2 | 42.979164 | 59.4312 | 12.831484 | 9.4958 |
| | | 0.4 | 30.407337 | 46.85 | 6.395548 | 5.182 |
| | | 0.6 | 13.890071 | 27.5746 | 1.7438921 | 1.9854 |
| | | 0.8 | 7.0071774 | 16.4269 | 0.39057034 | 0.6433 |
| | | 1.0 | 4.316065 | 8.9748 | 0.09733833 | 0.2396 |
| | Coma | 0.05 | 44.961002 | 60.9391 | 14.367914 | 10.5027 |
| | | 0.1 | 44.716515 | 60.9168 | 14.23886 | 10.4378 |
| | | 0.15 | 44.437935 | 60.4672 | 14.055115 | 10.2015 |
| | | 0.2 | 43.85603 | 59.7167 | 13.669502 | 10.0244 |
| | | 0.4 | 35.743027 | 48.2754 | 9.471531 | 6.3634 |
| | | 0.6 | 22.165663 | 29.2766 | 3.681941 | 2.1704 |
| | | 0.8 | 14.670907 | 18.1658 | 1.2350944 | 1.1757 |
| | | 1.0 | 9.214631 | 10.4162 | 0.2814782 | 0.9973 |
| | Spherical | 0.05 | 45.054314 | 60.9665 | 14.31999 | 10.516 |
| | | 0.1 | 44.98696 | 60.8835 | 14.201583 | 10.416 |
| | | 0.15 | 44.794838 | 60.7608 | 14.022252 | 10.152 |
| | | 0.2 | 44.432343 | 59.9632 | 13.592614 | 9.726 |
| | | 0.4 | 34.84883 | 45.5462 | 7.6500244 | 4.6625 |
| | | 0.6 | 11.067788 | 15.912 | 0.91046876 | 0.982 |
| | | 0.8 | 10.69451 | 11.5452 | 0.8506638 | 0.8162 |
| | | 1.0 | 10.055914 | 9.6316 | 1.0146351 | 1.1397 |
| | Trefoil | 0.05 | 45.060875 | 60.9413 | 14.3658 | 10.4994 |
| | | 0.1 | 44.894955 | 60.8259 | 14.229303 | 10.4150 |
| | | 0.15 | 44.516296 | 60.5646 | 13.958146 | 10.0474 |
| | | 0.2 | 43.94043 | 59.8601 | 13.487194 | 9.5037 |
| | | 0.4 | 36.660576 | 50.0387 | 8.857303 | 5.4211 |
| | | 0.6 | 23.58351 | 34.8188 | 4.0978146 | 2.7933 |
| | | 0.8 | 16.700645 | 24.4844 | 2.4396265 | 1.3593 |
| | | 1.0 | 16.576807 | 18.3695 | 2.3931422 | 1.1536 |

**Table S8:** AP50 and AP75 value comparison of Cellpose2.0 and anchor-free model under mixed optical aberrations.

| | | AP50 | | AP75 | |
|---|---|---|---|---|---|
| | | Cellpose2.0 | anchor-free | Cellpose2.0 | anchor-free |
| A172 | 4-6 | 10.12826 | 43.1541 | 3.7755587 | 5.3736 |
| | 4-8 | 12.933419 | 43.1498 | 4.6606636 | 6.1091 |
| | 4-13 | 12.677252 | 41.6636 | 4.6677685 | 5.7252 |
| | 4-18 | 6.0376587 | 42.0280 | 1.9841688 | 4.5092 |
| BT474 | 4-6 | 17.456465 | 50.1168 | 11.723929 | 15.8447 |
| | 4-8 | 18.542797 | 53.3629 | 12.8786545 | 17.5923 |
| | 4-13 | 19.13368 | 53.4803 | 13.15696 | 17.5013 |
| | 4-18 | 14.310344 | 43.1462 | 8.771276 | 12.2371 |
| BV2 | 4-6 | 51.3882 | 79.5191 | 31.344627 | 16.1079 |
| | 4-8 | 54.313477 | 80.7567 | 34.610855 | 19.2611 |
| | 4-13 | 54.837345 | 80.8237 | 33.571056 | 18.8267 |
| | 4-18 | 45.41742 | 77.6225 | 25.813229 | 10.5782 |
| Huh7 | 4-6 | 2.2506728 | 20.4631 | 0.915206 | 2.4742 |
| | 4-8 | 2.378491 | 22.1514 | 0.9905247 | 2.7686 |
| | 4-13 | 2.3541937 | 22.9800 | 0.88813007 | 2.6120 |
| | 4-18 | 1.61945 | 15.6167 | 0.6233179 | 1.7741 |
| MCF7 | 4-6 | 21.736025 | 54.9959 | 11.390692 | 11.0783 |
| | 4-8 | 24.973265 | 58.8781 | 13.512619 | 13.0116 |
| | 4-13 | 25.763735 | 60.1113 | 13.429858 | 13.3193 |
| | 4-18 | 14.965471 | 47.1988 | 7.3563566 | 7.6731 |
| SkBr3 | 4-6 | 78.33067 | 83.6065 | 53.432644 | 26.7937 |
| | 4-8 | 78.68467 | 83.8679 | 55.52305 | 28.4495 |
| | 4-13 | 78.80779 | 84.3040 | 54.608196 | 28.0465 |
| | 4-18 | 75.90009 | 82.1667 | 51.53406 | 23.5138 |
| SKOV3 | 4-6 | 7.1198688 | 48.9834 | 3.0168588 | 7.2245 |
| | 4-8 | 8.551794 | 51.9416 | 3.3482785 | 8.3811 |
| | 4-13 | 8.505803 | 51.6403 | 3.243553 | 8.2761 |
| | 4-18 | 4.892781 | 40.0999 | 2.1806016 | 4.6624 |
| SHSY5Y | 4-6 | 8.909224 | 29.4983 | 2.2686622 | 2.6636 |
| | 4-8 | 10.889745 | 30.4593 | 2.6795132 | 2.6121 |
| | 4-13 | 11.892 | 30.1896 | 2.73786 | 2.4669 |
| | 4-18 | 3.871907 | 25.3608 | 0.9635775 | 1.7815 |